\documentclass[11pt]{article}

\usepackage[left=1.0in,right=1.0in,top=1.0in,bottom=1.0in]{geometry}
\usepackage{float}
\usepackage{xcolor}
\usepackage{bbm}
\usepackage{amsmath}
\usepackage{caption}
\usepackage{graphicx}
\usepackage{authblk}
\usepackage{titlesec}
\usepackage{url}
\usepackage{natbib}
\usepackage[flushmargin]{footmisc}
\usepackage{setspace}
\usepackage{cleveref}

\providecommand{\keywords}[1]{\textbf{\textbf{Keywords:}} #1}
\def\spacingset#1{\renewcommand{\baselinestretch}%
{#1}\small\normalsize} \spacingset{1}

\begin{document}

  \title{Measuring Spatial Allocative Efficiency in Basketball}
  
  \author{Nathan Sandholtz}
  \author{Jacob Mortensen}
  \author{Luke Bornn}
 
 \affil{Simon Fraser University}

\maketitle

  \begin{abstract}
Every shot in basketball has an opportunity cost; one player's shot eliminates all potential opportunities from their teammates for that play.  For this reason, player-shot efficiency should ultimately be considered relative to the lineup.  This aspect of efficiency---the optimal way to allocate shots within a lineup---is the focus of our paper.  Allocative efficiency should be considered in a spatial context since the distribution of shot attempts within a lineup is highly dependent on court location.  We propose a new metric for spatial allocative efficiency by comparing a player's field goal percentage (FG\%) to their field goal attempt (FGA) rate in context of both their four teammates on the court and the spatial distribution of their shots.  Leveraging publicly available data provided by the National Basketball Association (NBA), we estimate player FG\% at every location in the offensive half court using a Bayesian hierarchical model.  Then, by ordering a lineup's estimated FG\%s and pairing these rankings with the lineup's empirical FGA rate rankings, we detect areas where the lineup exhibits inefficient shot allocation.  Lastly, we analyze the impact that sub-optimal shot allocation has on a team's overall offensive potential, demonstrating that inefficient shot allocation correlates with reduced scoring.
\end{abstract}
    \\
   \medskip \\
\textit{*The first and second authors contributed equally to this work.}

\spacingset{1.2} 

\section{Introduction} 

From 2017 to 2019, the Oklahoma City Thunder faced four elimination games across three playoff series.  In each of these games, Russell Westbrook attempted over 30 shots and had an average usage rate of 45.5\%.\footnote{Usage percentage is an estimate of the percentage of team plays used by a player while they were on the floor.  For a detailed formula see \url{www.basketball-reference.com/about/glossary.html}}  The game in which Westbrook took the most shots came in the first round of the 2017-18 National Basketball Association (NBA) playoffs, where he scored 46 points on 43 shot attempts in a 96-91 loss to the Utah Jazz.  At the time, many popular media figures conjectured that having one player dominate field goal attempts in this way would limit the Thunder's success.  While scoring 46 points in a playoff basketball game is an impressive feat for any one player, its impact on the overall game score is moderated by the fact that it required 43 attempts.  Perhaps not coincidentally, the Thunder lost three of these four close-out games and never managed to make it out of the first round of the playoffs.  

At its core, this critique is about shot efficiency.  The term `shot efficiency' is used in various contexts within the basketball analytics community, but in most cases it has some reference to the average number of points a team or player scores per shot attempt.  Modern discussion around shot efficiency in the NBA typically focuses on either shot selection or individual player efficiency.  The concept of shot selection efficiency is simple: 3-pointers and shots near the rim have the highest expected points per shot, so teams should prioritize these high-value shots.  The idea underlying individual player efficiency is also straightforward; scoring more points on the same number of shot attempts increases a team's overall offensive potential.  

However, when discussing a player's individual efficiency it is critical to do so in context of the lineup.  Basketball is not a 1-v-1 game, but a 5-v-5 game.  Therefore, when a player takes a shot, the opportunity cost not only includes all other shots this player could have taken later in the possession, but also the potential shots of their four teammates.  So regardless of a player's shooting statistics relative to the league at large, a certain dimension of shot efficiency can only be defined relative to the abilities of a player's teammates.  Applying this to the Oklahoma City Thunder example above, if Westbrook were surrounded by dismal shooters, 43 shot attempts might not only be defensible but also desirable.  On the other hand, if his inordinate number of attempts prevented highly efficient shot opportunities from his teammates, then he caused shots to be inefficiently distributed and decreased his team's scoring potential.  This aspect of efficiency---the optimal way to allocate shots within a lineup---is the primary focus of our paper. 

\begin{figure}[H]
 \begin{center}
\includegraphics[trim={0 0cm 0 0cm}, clip, width=.49\textwidth]{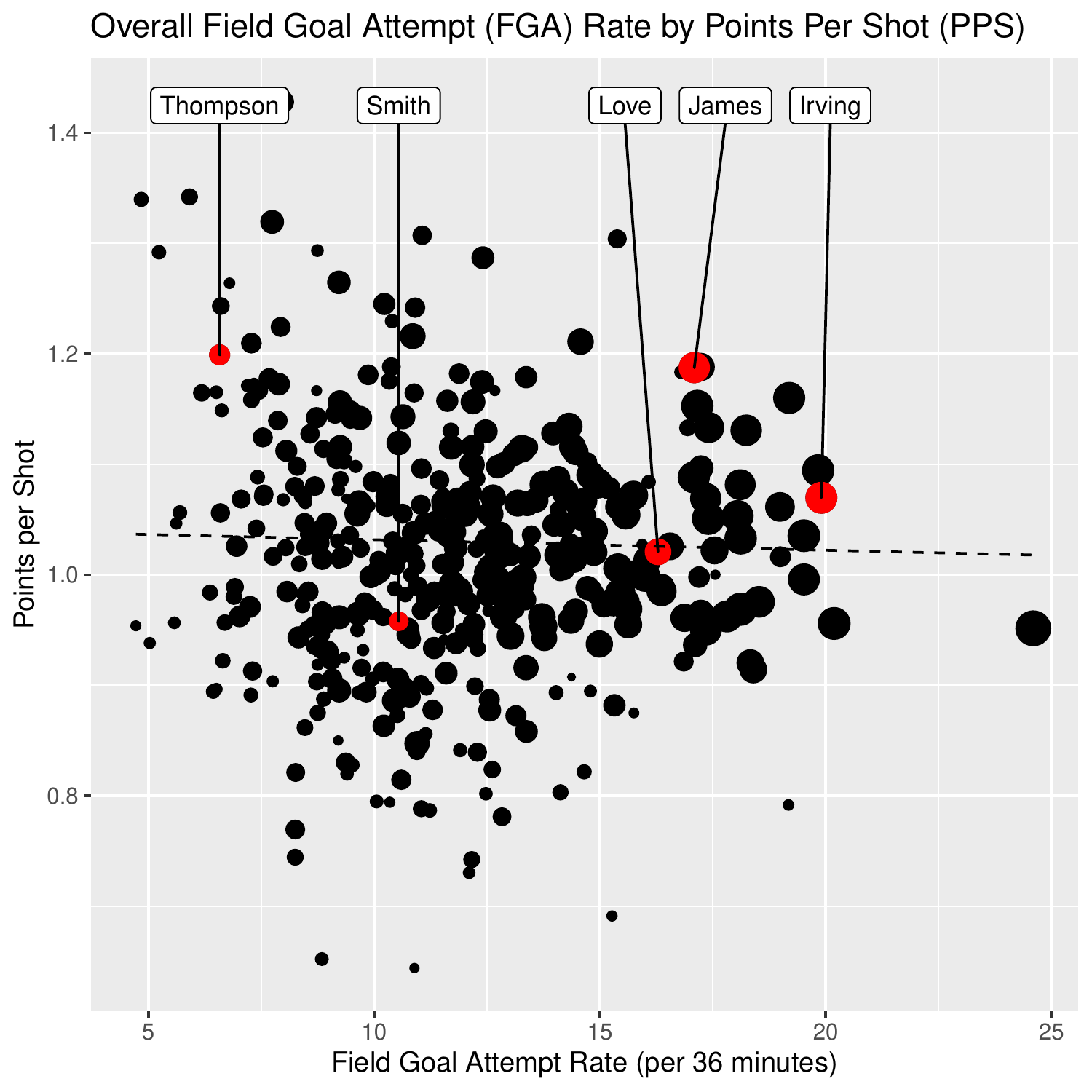}
\includegraphics[trim={0 0cm 0 0cm}, clip, width=.49\textwidth]{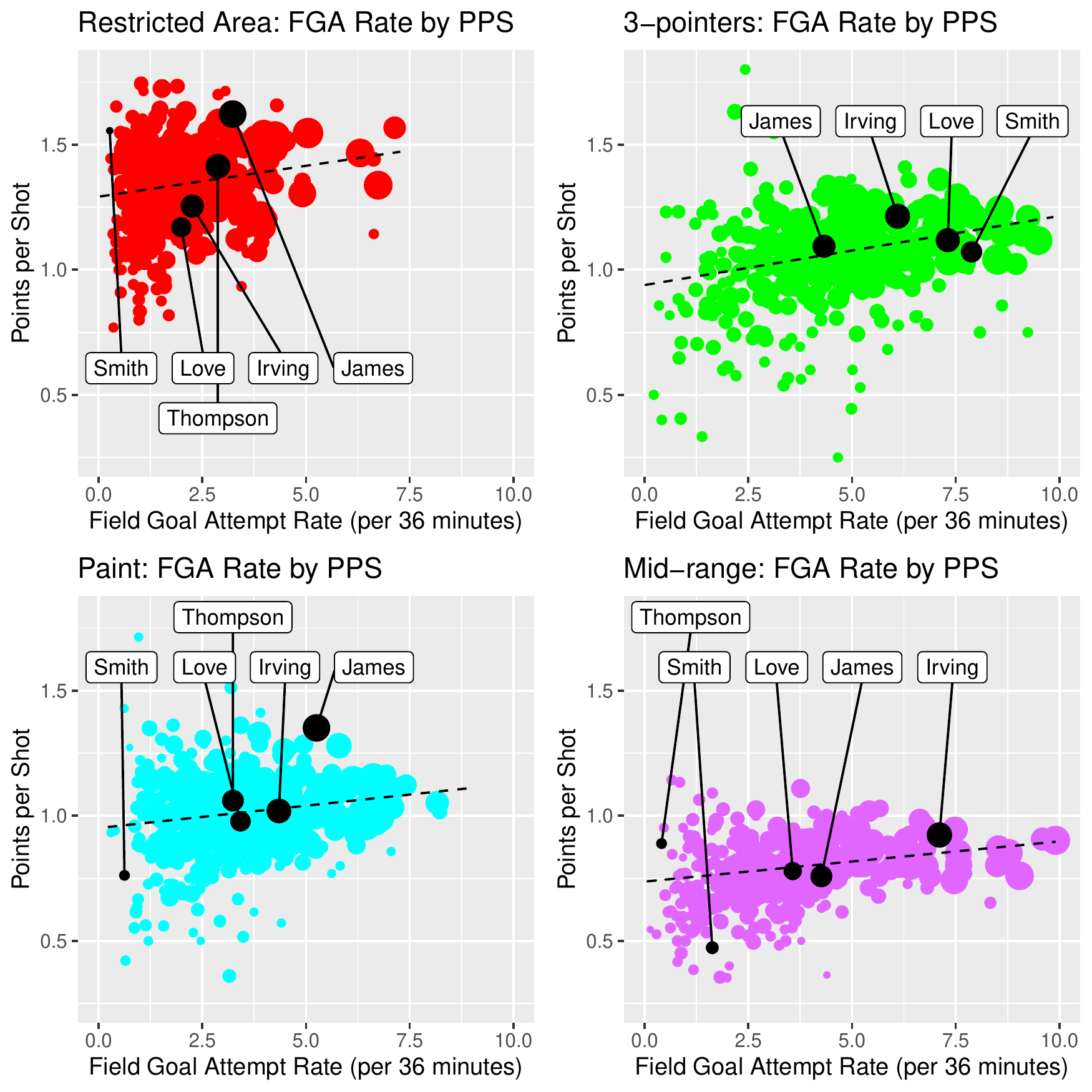}
\end{center}
\caption{Left: overall relationship between field goal attempt rate (x-axis) and points per shot (y-axis). Right: same relationship conditioned on various court regions.  The Cleveland Cavaliers 2016-17 starting lineup is highlighted in each plot.  The weighted least squares fit of each scatter plot is overlaid in each plot by a dotted line.}
\label{fig:simpsons}
\end{figure}

Allocative efficiency is spatially dependent.  As illustrated in Figure \ref{fig:simpsons}, the distribution of shots within a lineup is highly dependent on court location.  The left plot in Figure \ref{fig:simpsons} shows the overall relationship between shooting frequency (x-axis) and shooting skill (y-axis), while the four plots on the right show the same relationship conditioned on various court regions.  Each dot represents a player, and the size of the dot is proportional to the number of shots the player took over the 2016-17 NBA regular season.  To emphasize how shot allocation within lineups is spatially dependent, we have highlighted the Cleveland Cavaliers starting lineup, consisting of LeBron James, Kevin Love, Kyrie Irving, JR Smith, and Tristan Thompson. 

When viewing field goal attempts without respect to court location (left plot), Kyrie Irving appears to shoot more frequently than both Tristan Thompson and LeBron James, despite scoring fewer points per shot than either of them. However, after conditioning on court region (right plots), we see that Irving only has the highest FGA rate in the mid-range region, which is the region for which he has the highest PPS for this lineup.  James takes the most shots in the restricted area and paint regions---regions in which he is the most efficient scorer.  Furthermore, we see that Thompson's high overall PPS is driven primarily by his scoring efficiency from the restricted area and that he has few shot attempts outside this area. Clearly, understanding how to efficiently distribute shots within a lineup must be contextualized by spatial information. 

 Notice that in the left panel of Figure \ref{fig:simpsons}, the relationship between field goal attempt (FGA) rate and points per shot (PPS) appears to be slightly negative, if there exists a relationship at all. Once the relationship between FGA rate and PPS is spatially disaggregated (see right hand plots of Figure \ref{fig:simpsons}), the previously negative relationship between these variables becomes positive in every region. This instance of Simpson's paradox has non-trivial implications in the context of allocative efficiency which we will discuss in the following section. 
 
The goal of our project is to create a framework to assess the strength of the relationship between shooting frequency and shooting skill spatially within lineups and to quantify the consequential impact on offensive production.  Using novel metrics we develop, we quantify how many points are being lost through inefficient spatial lineup shot allocation, visualize where they are being lost, and identify which players are responsible.  

\subsection{Related Work} \label{sec:related_work}

In recent years, a number of metrics have been developed which aim to measure shot efficiency, such as true shooting percentage \citep{kubatko2007}, qSQ, and qSI \citep{chang2014}.  Additionally, metrics have been developed to quantify individual player efficiency, such as Hollinger's player efficiency rating \citep{bbref_per}.  While these metrics intrinsically account for team context, there have been relatively few studies which have looked at shooting decisions explicitly in context of lineup, and none spatially.

\cite{goldman2011} coined the term `allocative efficiency', modeling the decision to shoot as a dynamic mixed-strategy equilibrium weighing both the continuation value of a possession and the value of a teammate's potential shot.  They propose that a team achieves optimal allocative efficiency when, at any given time, the lineup cannot reallocate the ball to increase productivity on the margin.  Essentially, they argue that lineups optimize over all dimensions of an offensive strategy to achieve equal marginal efficiency for every shot.  The left plot of Figure \ref{fig:simpsons} is harmonious with this theory---there appears to be no relationship between player shooting frequency and player shooting skill when viewed on the aggregate.  However, one of the most important dimensions the players optimize over is court location.  Once we disaggregate the data by court location, (as shown in the right plots of Figure \ref{fig:simpsons}), we see a clear relationship between shooting frequency and shooting skill.  A unique contribution of our work is a framework to assess this spatial component of allocative efficiency. 

‘Shot satisfaction’ \citep{cervone2016} is another rare example of a shot efficiency metric that considers lineups.  Shot satisfaction is defined as the expected value of a possession conditional on a shot attempt (accounting for various contextual features such as the shot location, shooter, and defensive pressure at the time of the shot) minus the unconditional expected value of the play.  However, since shot satisfaction is marginalized over the allocative and spatial components, these factors cannot be analyzed using this metric alone.  Additionally, shot satisfaction is dependent on proprietary data which limits its availability to a broad audience. 

\subsection{Data and Code} 

The data used for this project is publicly available from the NBA stats API (stats.nba.com).  Shooter information and shot $(x, y)$ locations are available through the ‘shotchartdetail' API endpoint, while lineup information can be constructed from the ‘playbyplayv2' endpoint.  Code for constructing lineup information from play-by-play data is available at: \url{https://github.com/jwmortensen/pbp2lineup}.  Using this code, we gathered a set of 224,567 shots taken by 433 players during the 2016-17 NBA regular season, which is the data used in this analysis.  Code used to perform an empirical version of the analysis presented in this paper is also available online: \url{https://github.com/nsandholtz/lpl}.

\section{Models} 

The foundation of our proposed allocative efficiency metrics rest on spatial estimates of both player FG\% and field goal attempt (FGA) rates.  With some minor adjustments, we implement the FG\% model proposed in \cite{cervone2016}.  As this model is the backbone of the metrics we propose in Section 3, we thoroughly detail the components of their model in Section 2.1. In Section 2.2, we present our model for estimating spatial FGA rates.

\subsection{Estimating FG\% Surfaces} 

 Player FG\% is a highly irregular latent quantity over the court space.  In general, players make more shots the closer they are to the hoop, but some players are more skilled from a certain side of the court and others specialize from very specific areas, such as the corner 3-pointer.  In order to capture these kinds of non-linear relationships, \cite{cervone2016} summarizes the spatial variation in player shooting skill by a Gaussian process represented by a low-dimensional set of deterministic basis functions.  Player-specific weights are estimated for the basis functions using a Bayesian hierarchical model \citep{gelman2013bayesian}.  This allows the model to capture nuanced spatial features that player FG\% surfaces tend to exhibit, while maintaining a feasible dimensionality for computation.

We model the logit of $\pi_j(\mathbf{s})$, the probability that player $j$ makes a shot at location $\mathbf{s}$, as a linear model:
\begin{align}
    \text{log}\Big(\frac{\pi_j(\mathbf{s})}{1 - \pi_j(\mathbf{s})}\Big) = \pmb{\beta}^\prime \mathbf{x} + Z_j(\mathbf{s}). %[\mathbf{w}_{j}]'~~~\pmb{\Lambda}~~~\pmb{\Psi}(\mathbf{z})} + \epsilon
    \label{eq:main_model}
\end{align}
 Here $\pmb{\beta}$ is a $4 \times 1$ vector of covariate effects and $\mathbf{x}$ is a $4 \times 1$ vector of observed covariates for the shot containing an intercept, player position, shot distance, and the interaction of player position and shot distance. $Z_j(\mathbf{s})$ is a Gaussian process which accounts for the impact of location on the probability of player $j$ making a shot and is modeled using a functional basis representation,
 \begin{equation}
    Z_j(\mathbf{s}) = \mathbf{w}_j^\prime \pmb{\Lambda} \pmb{\Psi}(\mathbf{s}),
    \label{eq:gp_construction}
 \end{equation}
 where $\mathbf{w}_j = (\text{w}_{j1}, \dots, \text{w}_{jD})^\prime$ denotes the latent basis function weights for player $j$ and $\pmb{\Lambda} \pmb{\Psi}(\mathbf{s})$ denotes the basis functions.  Specifically, $\pmb{\Lambda} = (\pmb{\lambda}_1^\prime, \dots, \pmb{\lambda}_D^\prime)^\prime$ is a $D \times K$ matrix, where each row vector $\pmb{\lambda}_d$ represents the projection of the $d$th basis function onto a triangular mesh with $K$ vertices over the offensive half court (more details on the construction of $\pmb{\Lambda}$ follow below). We use the mesh proposed in \cite{cervone2016}, which was selected specifically for modeling offensive spatial behaviour in basketball. $\pmb{\Psi}(\mathbf{s}) = (\psi_1(\mathbf{s}),\dots,\psi_K(\mathbf{s}))^\prime$ is itself a vector of basis functions where each $\psi_k(\mathbf{s})$ is 1 at mesh vertex $k$, 0 at all other vertices, and values at the interior points of each triangle are determined by linear interpolation between vertices (see \cite{Lindgren2011} for details).  Finally, we assume $\mathbf{w}_j \sim \mathcal{N}(\pmb{\omega}_j, \pmb{\Sigma}_j)$, which makes (\ref{eq:gp_construction}) a Gaussian process with mean $\pmb{\omega}_j^\prime \pmb{\Lambda} \pmb{\Psi}(\mathbf{s})$ and covariance function Cov$(\mathbf{s}_1, \mathbf{s}_2) = \pmb{ \Psi}(\mathbf{s}_1)^\prime \pmb{\Lambda}^\prime \pmb{\Sigma}_j \pmb{\Lambda} \pmb{\Psi}(\mathbf{s}_2)$.
  
\begin{figure}[H]
 \begin{center}
\includegraphics[trim={0 7cm 0 7cm}, clip, width=1\textwidth]{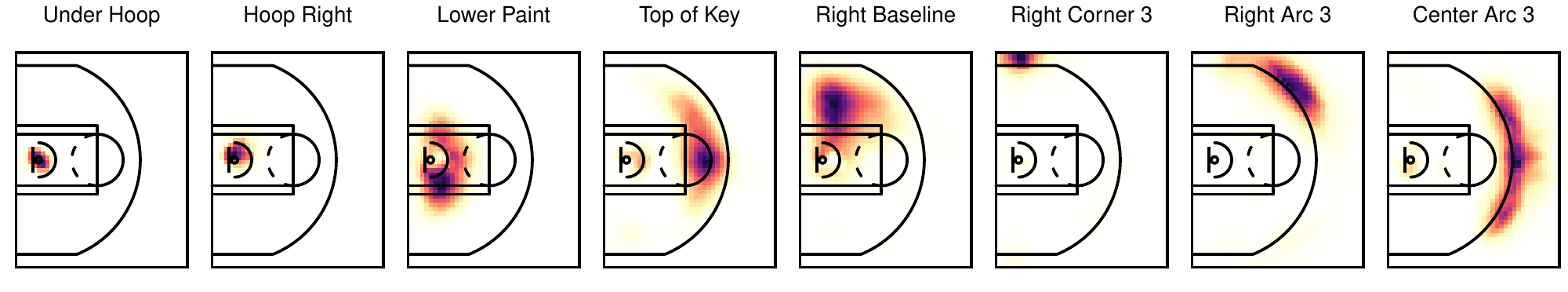}
\includegraphics[trim={0 7cm 0 7cm}, clip, width=1\textwidth]{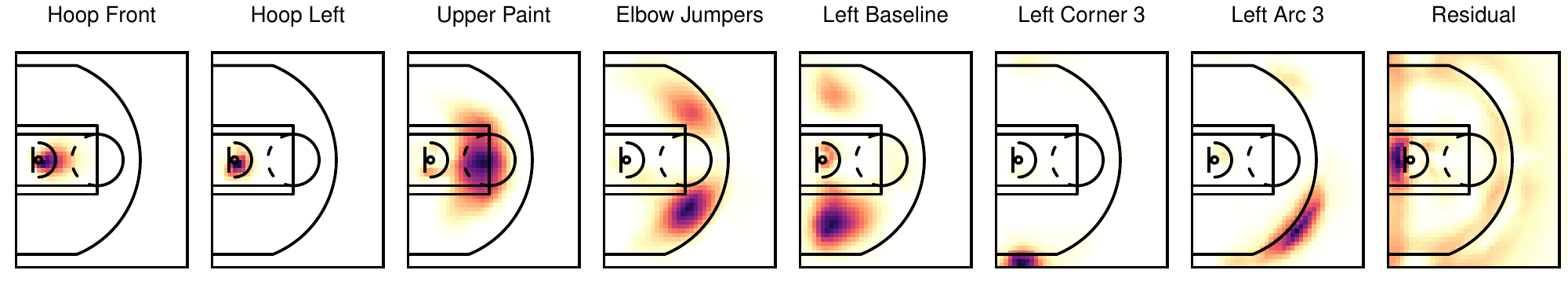}
\end{center}
\caption{Deterministic bases resulting from the non-negative matrix factorization of $\mathbf{P}$.  The plots are arranged such that the bases closest to the hoop are on the left (e.g. Under Hoop) and the bases furthest from the hoop are on the right (e.g. Center Arc 3).  The residual basis, comprising court locations where shots are infrequently attempted from, is shown in the bottom-right plot.}
\label{fig:Lambda}
\end{figure}

Following \cite{Miller2014}, the bases of shot taking behavior, $\pmb \Lambda$, are computed through a combination of smoothing and non-negative matrix factorization (NMF) \citep{Lee1999}.  Using integrated nested Laplace approximation (INLA) as the engine for our inference, we first fit a log Gaussian Cox Process (LGCP) \citep{Banerjee2015} independently to each player's point process defined by the $(x,y)$ locations of their made shots using the aforementioned mesh.\footnote{Players who took less than five shots in the regular season are treated as  ``replacement players.''}  Each player's estimated intensity function is evaluated at each vertex, producing a $K$-dimensional vector for each of the $L = 433$ players in our data.  These vectors are exponentiated and gathered (by rows) into the $L \times K$ matrix $\mathbf{P}$, which we then factorize via NMF:
 \begin{align}
     \mathbf{P} \approx \bigg(\underset{L\times D}{\mathbf{B}}\bigg) \bigg(\underset{D\times K}{\pmb{\Lambda} }\bigg). \label{eq:NMF}
 \end{align}
 This yields $\pmb \Lambda$, the deterministic bases we use in \eqref{eq:gp_construction}.  While the bases from (\ref{eq:NMF}) are constructed solely with respect to the spatial variation in the FGA data (i.e. no basketball-specific structures are induced a priori), the constraint on the number of bases significantly impacts the basis shapes.  In general, the NMF tends to first generate bases according to shot distance.  After accounting for this primary source of variation, other systematic features of variation begin to appear in the bases, notably asymmetry.   We use D = 16 basis functions, aligning with \cite{Miller2014} which suggests the optimal number of basis functions falls between 15 and 20.  Collectively, these bases comprise a comprehensive set of shooting tendencies, as shown in Figure \ref{fig:Lambda}.  We have added labels post hoc to provide contextual intuition.
 
Conceptually, the $Z_j(\mathbf{s})$ term in (\ref{eq:main_model}) represents a player-specific spatial `correction' to the global regression model $\pmb{\beta}^\prime \mathbf{x}$.  These player-specific surfaces are linear combinations of the bases shown in Figure \ref{fig:Lambda}.  The weights of these combinations, $\mathbf{w}_j$, are latent parameters which are jointly estimated with $\pmb{\beta}$.  Since these player weights can be highly sensitive for players with very little data, it is imperative to introduce a regularization mechanism on them, which is accomplished using a conditionally autoregressive (CAR) prior.  Conveniently, the NMF in (\ref{eq:NMF}) provides player-specific loadings onto these bases, $\mathbf{B}$, which we use in constructing this CAR prior on the basis weights, $\mathbf{w}_j$ \citep{Besag1974}.  The purpose of using a CAR prior on the basis weights is to shrink the FG\% estimates of players with similar shooting characteristics toward each other.  This is integral for obtaining realistic FG\% estimates in areas where a player took a low volume of shots.  With only a handful of shots from an area, a player's empirical FG\% can often be extreme (e.g. near 0\% or 100\%).  The CAR prior helps to regularize these extremes by borrowing strength from the player's neighbors in the estimation.  
 
 In order to get some notion of shooting similarity between players, we calculate the Euclidean distance between the player loadings contained in $\mathbf{B}$ and, for a given player, define the five players with the closest player loadings as their neighbors.  This is intentionally chosen to be fewer than the number of neighbors selected by Cervone, recognizing that more neighbors defines a stronger prior and limits player-to-player variation in the FG\% surfaces. We enforce symmetry in the nearest-neighbors relationship by assuming that if player $j$ is a neighbor of player $\ell$, then player $\ell$ is also a neighbor of player $j$, which results in some players having more than five neighbors.  These relationships are encoded in a player adjacency matrix $\mathbf{H}$ where entry $(j, \ell)$ is 1 if player $\ell$ is a neighbor of player $j$ and 0 otherwise.  The CAR prior on $\mathbf{w}_{j}$ can be specified as
\begin{align}
    (\mathbf{w}_{j} | \mathbf{w}_{-(j)}, \tau^2) &\sim \mathcal{N}\Bigg(\frac{1}{n_j}\sum_{\ell:H_{j\ell} = 1} \mathbf{w}_{\ell}, \frac{\tau^2}{n_j}\mathbf{I}_D \Bigg) \\
    \tau^2 &\sim \text{InvGam}(1,1). \nonumber
\end{align}
where $n_j$ is the total number of neighbors for player $j$.  

\begin{figure}[H]
\includegraphics[trim={.5cm 5cm 0cm 5cm}, clip, width=1\textwidth]{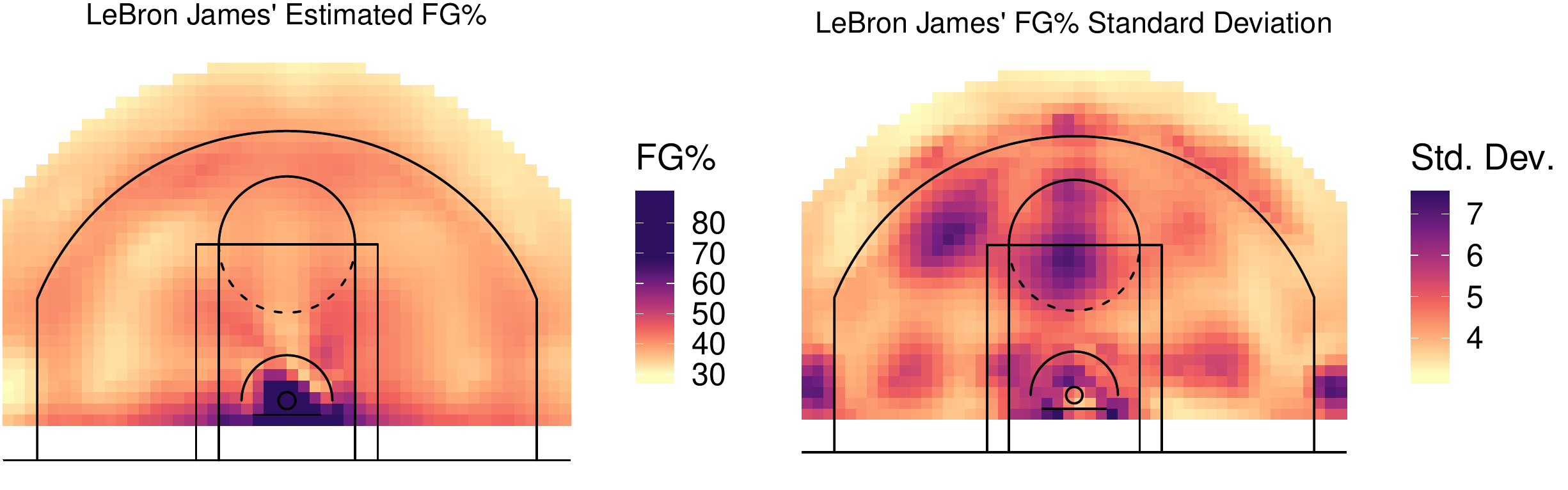}
\caption{LeBron James 2016-17 FG\% posterior mean (left) and posterior standard deviation (right) projected onto the offensive half court.  The prediction surfaces shown here and throughout the figures in this paper utilize projections onto a spatial grid of 1 ft. by 1 ft. cells.}
\label{fig:example_fgp_surf}
\end{figure}

Lastly, we set a $\mathcal{N}(\mathbf{0}, 0.001 \times \mathbf{I})$ prior on $\pmb{\beta}$, and fit the model using INLA.  This yields a model that varies spatially and allows us to predict player-specific FG\% at any location in the offensive half court. In order to get high resolution FG\% estimates, we partition the court into 1 ft. by 1 ft. grid cells (yielding a total of $M$ = 2350 cells) and denote player $j$'s FG\% at the centroid of grid cell $i$ as $\xi_{ij}$.  The projection of the FG\% posterior mean ($\widehat{\pmb{\xi}}_{j}$) for LeBron James is depicted in Figure \ref{fig:example_fgp_surf}.  

In order to have sufficient data to reliably estimate these surfaces, we assume that player FG\%s are lineup independent.  We recognize this assumption may be violated in some cases, as players who draw significant defensive attention can improve the FG\%s of their teammates by providing them with more unguarded shot opportunities.  Additionally, without defensive information about the shot opportunities, the FG\% estimates are subject to systematic bias.  Selection bias is introduced by unequal amounts of  defensive  pressure  applied  to  shooters  of  different  skill levels.

The Bayesian modeling framework can amplify selection bias as well.  Since the FG\% estimates are regularized in our model via a CAR prior,  players FG\% estimates shrink toward their neighbors (which we've defined in terms of FGA rate).  While this feature stabilizes estimates for players with low sample sizes, it can be problematic when entire neighborhoods have low sample sizes from specific regions.  For example, there are many centers who rarely or never shoot from long range.  Consequently, the entire neighborhood shrinks toward the global mean 3-point FG\%, inaccurately inflating these players' FG\%s beyond the 3-point line.  These are intriguing challenges and represent promising directions for future work. 

\subsection{Determining FGA Rate Surfaces} 

We determine a player's FGA rate surface by smoothing their shot attempts via a LGCP.  This model has the form $$\log{\lambda(\mathbf{s})} = \beta_0 + Z(\mathbf{s}),$$ where $\lambda(\mathbf{s})$ is the Poisson intensity indicating the number of expected shots at location $\mathbf{s}$, $\beta_0$ is an intercept, and $Z(\mathbf{s})$ is a Gaussian process. We fit this model separately for each player using INLA, following the approach in \cite{Simpson2015}.  In brief, they demonstrate that the likelihood for the LGCP can be approximated using a finite-dimensional Gaussian random field, allowing $Z(\mathbf{s})$ to be represented by the basis function expansion $Z(\mathbf{s}) = \sum_{b=1}^B z_b\phi_b(\mathbf{s})$. The basis function $\phi_b(\mathbf{s})$ projects shot location onto a triangular mesh akin to the one detailed for \eqref{eq:gp_construction}.  The expected value of $\lambda(\mathbf{s})$ integrated over the court is equal to the number of shots a player has taken, however there can be small discrepancies between the fitted intensity function and the observed number of shots.  In order to ensure consistency, we scale the resulting intensity function to exactly yield the player's observed number of shot attempts in that lineup. 

We normalize the surfaces to FGA per 36 minutes by dividing by the total number of minutes played by the associated lineup and multiplying by 36, allowing us to make meaningful comparisons between lineups who differ in the amount of minutes played.  As with the FG\% surfaces ($\pmb{\xi}$), we partition the court into 1 ft. by 1 ft. grid cells and denote player $j$'s FGA rate at the centroid of grid cell $i$ as $A_{ij}$.  

Note that we approach the FGA rate estimation from a fundamentally different perspective than the FG\% estimation.  We view a player's decision to shoot the ball as being completely within their control and hence non-random.  As such, we incorporate no uncertainty in the estimated surfaces.  We use the LGCP as a smoother for observed shots rather than as an estimate of a player's true latent FGA rate.  Other smoothing methods (e.g. kernel based methods \citep{Diggle1985}) could be used instead.

\begin{figure}[H]
 \centering
 \includegraphics[trim={0cm 2cm 0cm 1.75cm}, clip, width=1\textwidth]{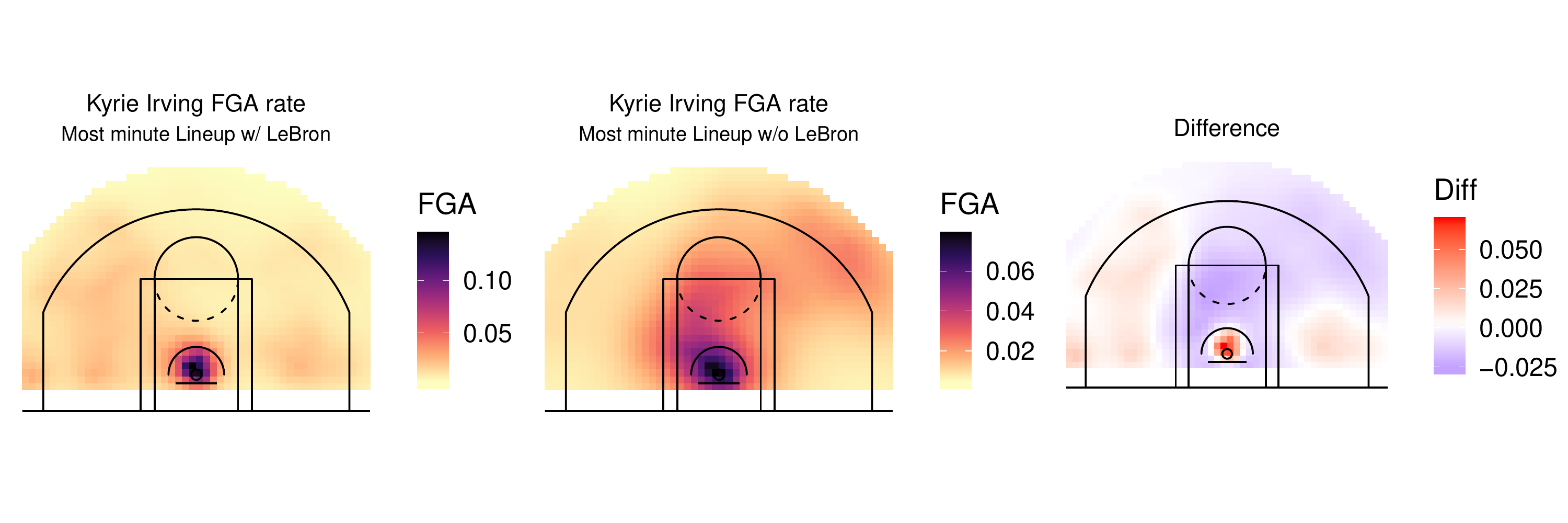}
\caption{Left: Kyrie Irving's FGA rate per 36 minutes in the starting lineup (in which he shared the most minutes with LeBron James).  Center: Kyrie Irving's FGA rate per 36 minutes in the lineup for which he played the most minutes without LeBron James.  Right: The difference of the center surface from the left surface.}
\label{fig:kyrie_fga_differences}
\end{figure}

Depending on the player and lineup, a player's shot attempt profile can vary drastically from lineup to lineup.  Figure \ref{fig:kyrie_fga_differences} shows Kyrie Irving's estimated FGA rate surfaces in the starting lineup (left) and the lineup in which he played the most minutes without LeBron James (middle).  Based on these two lineups, Irving took 9.2 more shots per 36 minutes when he didn't share the court with James.  He also favored the left side of the court far more, which James tends to dominate when on the court.  

Clearly player shot attempt rates are not invariant to their teammates on the court.  We therefore restrict player FGA rate estimation to lineup-specific data.  Fortunately, the additional sparsity introduced by conditioning on lineup is a non-issue.  If a player has no observed shot attempts from a certain region (e.g, Tristan Thompson from 3-point range), this simply means they chose not to shoot from that region---we don't need to borrow strength from neighboring players to shed light on this area of ``incomplete data".

\section{Allocative Efficiency Metrics}
\label{sec:lpl}

The models for FG\% and FGA rate described in Section 2 are the backbone of the allocative efficiency metrics we introduce in this section: lineup points lost (LPL) and player LPL contribution (PLC).  Before getting into the details, we emphasize that these metrics are agnostic to the underlying FG\% and FGA models; they can be implemented using even crude estimates of FG\% and FGA rate, for example, by dividing the court into discrete regions and using the empirical FG\% and FGA rate within each region.\footnote{Section \ref{sec:empirical_example} in  the  appendix  shows  how  LPL  can  be  calculated  using  empirical estimates of FG\% and FGA rate.  We use the Cavaliers starting lineup to compare these empirical LPL surfaces to the more sophisticated versions presented in the main text.}  Also note that the biases affecting FG\% and FGA rate described in Section 2 may affect the allocative efficiency metrics as well.  Section 4 includes a discussion of the causal limitations of the approach. 

LPL is the output of a two-step process.  First, we redistribute a lineup's observed distribution of shot attempts according to a proposed optimum.   This optimum is based on ranking the five players in the lineup with respect to their FG\% and FGA rate and then redistributing the shot attempts such that the FG\% ranks and FGA rate ranks match.  Second, we estimate how many points could have been gained had a lineup's collection of shot attempts been allocated according to this alternate distribution.  In this section, we go over each of these steps in detail and conclude by describing PLC, which measures how individual players contribute to LPL.    

\subsection{Spatial Rankings Within a Lineup}

With models for player FG\% and player-lineup FGA rate, we can rank the players in a given lineup (from 1 to 5) on these metrics at any spot on the court.  For a given lineup, let $\pmb{R}_{i}^{\xi}$ be a discrete transformation of $\pmb{\xi}_i$---the lineup's FG\% vector in court cell $i$---yielding each player's FG\% rank relative to their four teammates.  Formally,
 \begin{align}
     R_{ij}^{\xi} = \{(n_{{\xi_i}} + 1) - k : \xi_{ij} \equiv \xi^{(k)}_i\},
     \label{eq:fgp_rank}
 \end{align}
 where $n_{{\xi_i}}$ is the length of $\pmb{\xi}_i$, the vector being ranked (this length will always be 5 in our case), and $\xi^{(k)}_i$ is the $k$th order statistic of $\pmb{\xi}_i$.  Since $\xi_{ij}$ is a stochastic quantity governed by a posterior distribution, $R_{ij}^{\xi}$ is also distributional, however its distribution is discrete, the support being the integers $\{1,2,3,4,5\}$.  The distribution of $R_{ij}^{\xi}$ can be approximated by taking posterior samples of $\pmb{\xi}_i$ and ranking them via (\ref{eq:fgp_rank}).  Figure \ref{fig:example_fgp_ranks} in the appendix shows the 20\% quantiles, medians, and 80\% quantiles of the resulting transformed variates for the Cavaliers starting lineup.

\begin{figure}[H]
\includegraphics[trim={0cm 5cm 0cm 5.5cm}, clip, width=1\textwidth]{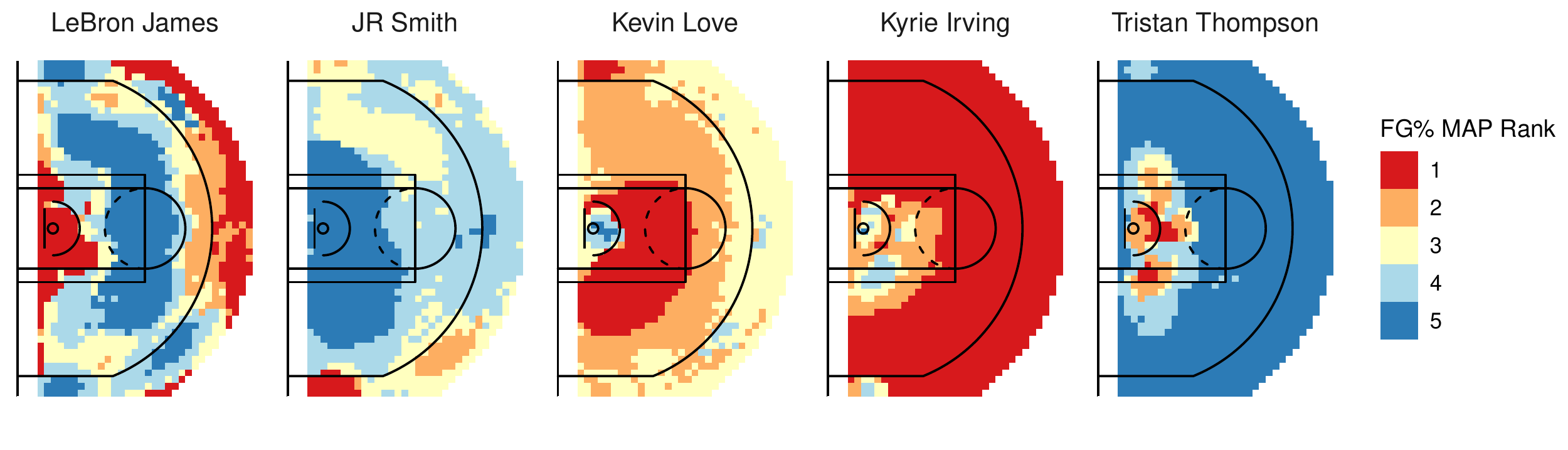}
\includegraphics[trim={0cm 1cm 0cm 11cm}, clip, width=1\textwidth]{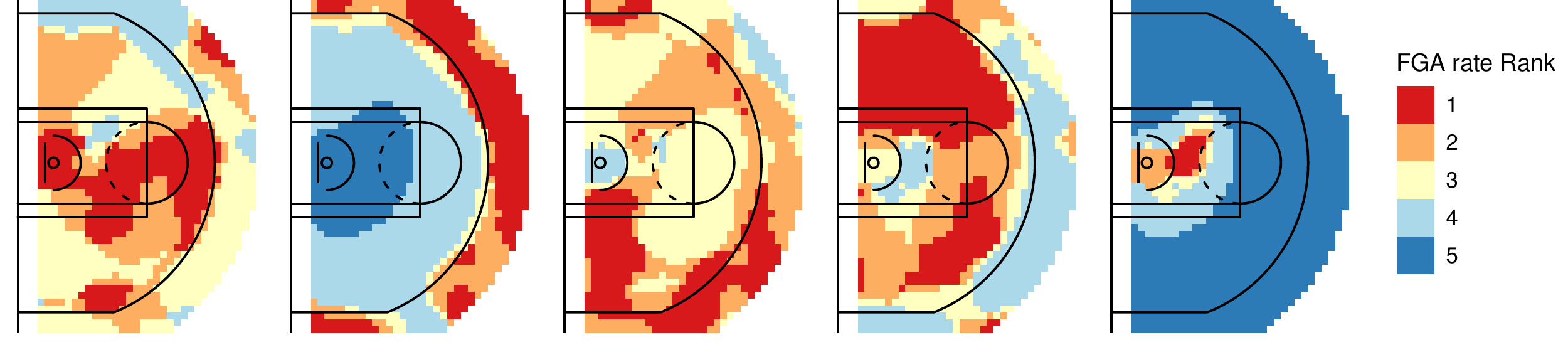}
\caption{Top: Estimated FG\% ranks for the Cleveland Cavaliers' starting lineup.  Bottom: Deterministic FGA rate ranks.}
\label{fig:example_fga_ranks}
\end{figure}

We obtain ranks for FGA rates in the same manner as for FG\%, except these will instead be deterministic quantities since the FGA rate surfaces, $\pmb{A}$, are fixed.  We define $R_{ij}^A$ as
 \begin{align}
     R_{ij}^{A} = \{(n_{{A_i}} + 1) - k : A_{ij} \equiv A^{(k)}_i\},
     \label{eq:fga_rank}
 \end{align}
where $n_{{A_i}}$ is the length of $\pmb{A}_i$ and $A^{(k)}_i$ is the $k$th order statistic of $\pmb{A}_i$.  Figure \ref{fig:example_fga_ranks} shows the estimated maximum a posteriori\footnote{For the FG\% rank surfaces we use the MAP estimate in order to ensure the estimates are always in the support of the transformation (i.e. to ensure $\widehat{R}_{ij}^{\xi} \in \{1,\ldots, 5\}$).  For parameters with continuous support, such as $\widehat{\pmb{\xi}}$, the hat symbol denotes the posterior mean.} (MAP) FG\% rank surfaces, $\widehat{\pmb{R}}^{\xi}$, and the deterministic FGA rate rank surfaces, $\pmb{R}^{A}$, for the Cleveland Cavaliers starting lineup. 

\begin{figure}[H]
\includegraphics[trim={0cm 4.5cm 0cm 4.4cm}, clip, width=1\textwidth]{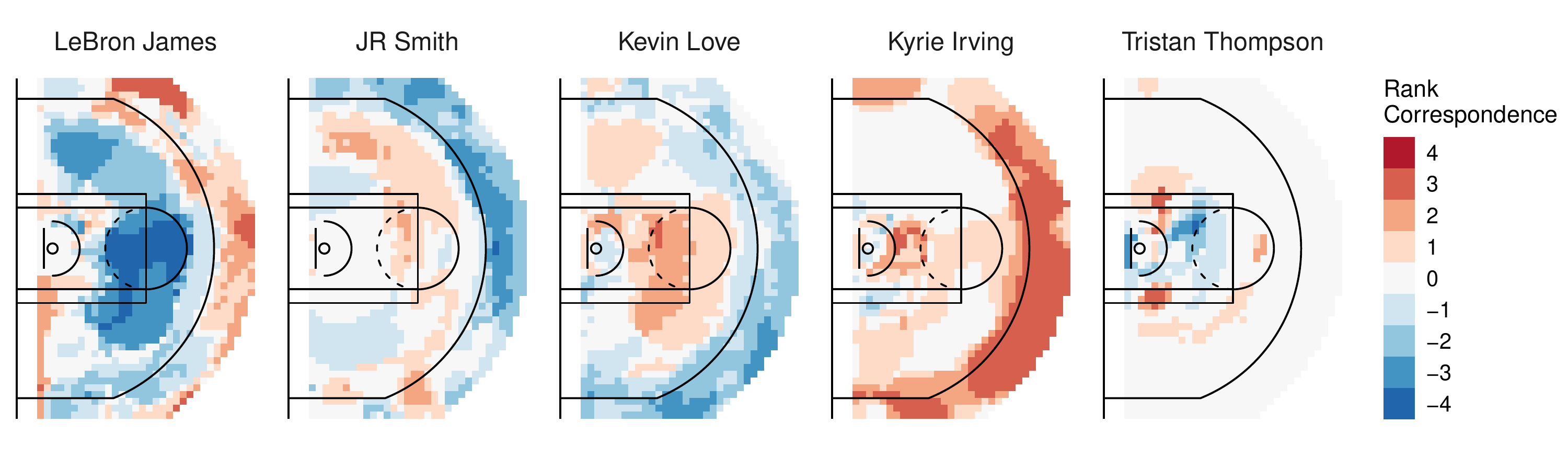}
\caption{Rank correspondence surfaces for the Cleveland Cavaliers' starting lineup.}
\label{fig:example_rank_corr}
\end{figure}

The strong correspondence between $\widehat{\pmb{R}}^{\xi}$ and $\pmb{R}^{A}$ shown in Figure \ref{fig:example_fga_ranks} is not surprising; all other factors being equal, teams would naturally want their most skilled shooters taking the most shots and the worst shooters taking the fewest shots in any given location.  By taking the difference of a lineup's FG\% rank surface from its FGA rate rank surface,  $\pmb{R}^{A} - \widehat{\pmb{R}}^{\xi}$, we obtain a surface which measures how closely the lineup's FG\% ranks match their FGA rate ranks.  Figure \ref{fig:example_rank_corr} shows these surfaces for the Cavaliers' starting lineup.  
 Note that rank correspondence ranges from -4 to 4.  A value of -4 means that the worst shooter in the lineup took the most shots from that location, while a positive 4 means the best shooter took the fewest shots from that location.  In general, positive values of rank correspondence mark areas of potential under-usage and negative values show potential over-usage.  For the Cavaliers, the positive values around the 3-point line for Kyrie Irving suggest that he may be under-utilized as a 3-point shooter.  On the other hand, the negative values for LeBron James in the mid-range region suggest that he may be over-used in this area.  We emphasize, however, that conclusions should be made carefully.  Though inequality between the FG\% and FGA ranks may be indicative of sub-optimal shot allocation, this interpretation may not hold in every situation due to bias introduced by confounding variables (e.g. defensive pressure, shot clock, etc.).

 \subsection{Lineup Points Lost}

 By reducing the FG\% and FGA estimates to ranks, we compromise the magnitude of player-to-player differences within lineups.   Here we introduce lineup points lost (LPL), which measures deviation from perfect rank correspondence while retaining the magnitudes of player-to-player differences in FG\% and FGA.  
 
 LPL is defined as the difference in expected points between a lineup's actual distribution of FG attempts, $\pmb{A}$, and a proposed redistribution, $\pmb{A}^*$, constructed to yield perfect rank correspondence (i.e. $\pmb{R}^{A^*} - \pmb{R}^{\xi} = \pmb{0}$).  Formally, we calculate LPL in the $i$th cell as 
  \begin{align}
  \text{LPL}_i &= \sum_{j = 1}^5 \text{v}_i \cdot \xi_{ij} \cdot \big(A_{i[g(R^{\xi}_{ij})]} - A_{ij}\big)  \label{eq:lpl1} \\
     &= \sum_{j = 1}^5 \text{v}_i \cdot \xi_{ij} \cdot \big(A^*_{ij} - A_{ij}\big), \label{eq:lpl2}
    % &= \sum_{j = 1}^5 (\text{v}_i \cdot \widehat{\xi}_{ij} \cdot A^*_{ij}) - (\text{v}_i \cdot \widehat{\xi}_{ij} \cdot A_{ij}) \\
   %  &= \sum_{j = 1}^5 \text{optimal}~ \mathbb{E} (\text{points}_{ij}) - \text{actual}~ \mathbb{E}(\text{points}_{ij})
    %  \label{eq:lpl3}
 \end{align}
 where $\text{v}_i$ is the point value (2 or 3) of a made shot, $\xi_{ij}$ is the FG\% for player $j$ in cell $i$, $A_{ij}$ is player $j$'s FG attempts (per 36 minutes) in cell $i$, and $g(R^{\xi}_{ij}) = \{k:~ R^{\xi}_{ij} \equiv R^A_{ik}\}$.  The function $g(\cdot)$ reallocates the observed shot attempt vector $\pmb{A}_{i}$ such that the best shooter always takes the most shots, the second best shooter takes the second most shots, and so forth. 
 
\begin{figure}[H]
\centering
\includegraphics[trim={0cm 0cm 0cm 0cm}, clip, width=1\textwidth]{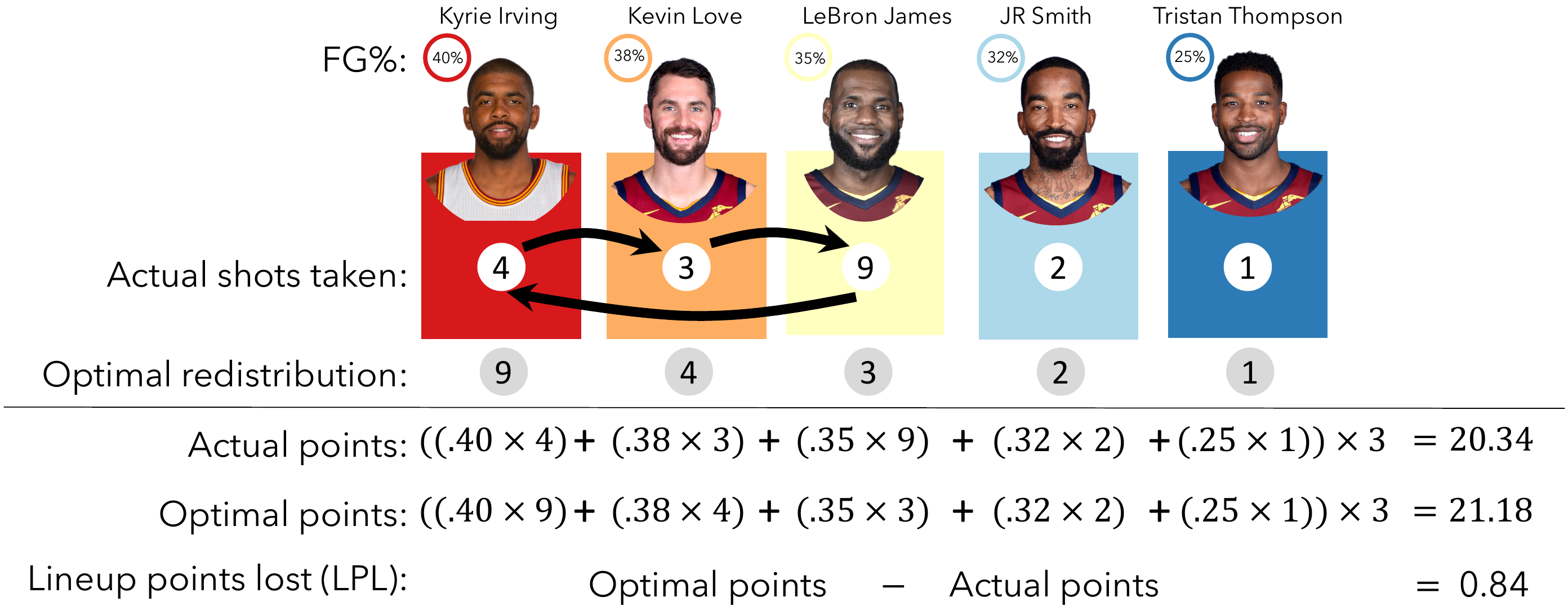}
\caption{A toy LPL computation in an arbitrary 3-point court region for the Cleveland Cavalier's starting lineup.  The players are ordered from left to right according to FG\% (best to worst).   Below each player's picture is the number of actual shots the player took from this location.  The black arrows show how the function $g(\cdot)$ reallocates these shots according to the players' FG\% ranks.  The filled gray dots show the number of shots the player would be allocated according to the proposed optimum.  Below the horizontal black line, each player's actual expected points and optimal expected points are calculated by multiplying their FG\% by the corresponding number of shots and the point value of the shot.  LPL is the difference (in expectation) between the optimal points and the actual points.}
\label{fig:toy_LPL}
\end{figure}

Figure \ref{fig:toy_LPL} shows a toy example of how LPL is computed for an arbitrary 3-point region, contextualized via the Cleveland Cavaliers starting lineup.  In this hypothetical scenario, James takes the most shots despite both Love and Irving being better shooters from this court region.  When calculating LPL for this region, Irving is allocated James' nine shots since he is the best shooter in this area. Love, as the second best shooter, is allocated Irving's four shots (which was the second most shots taken across the lineup).  James, as the third best shooter, is allocated the third most shot attempts (which is Love's three shots).  Smith and Thompson's shot allocations are unchanged since their actual number of shots harmonizes with the distribution imposed by $g(\cdot)$.  Each player's actual expected points and optimal expected points are calculated by multiplying their FG\% by the corresponding number of shots and the point-value of the shot (3 points in this case).  LPL is the difference (in expectation) between the optimal points and the actual points, which comes out to 0.84.

\begin{figure}[H]
\centering
\includegraphics[trim={0cm 2.5cm 0cm 2.4cm}, clip, width=1\textwidth]{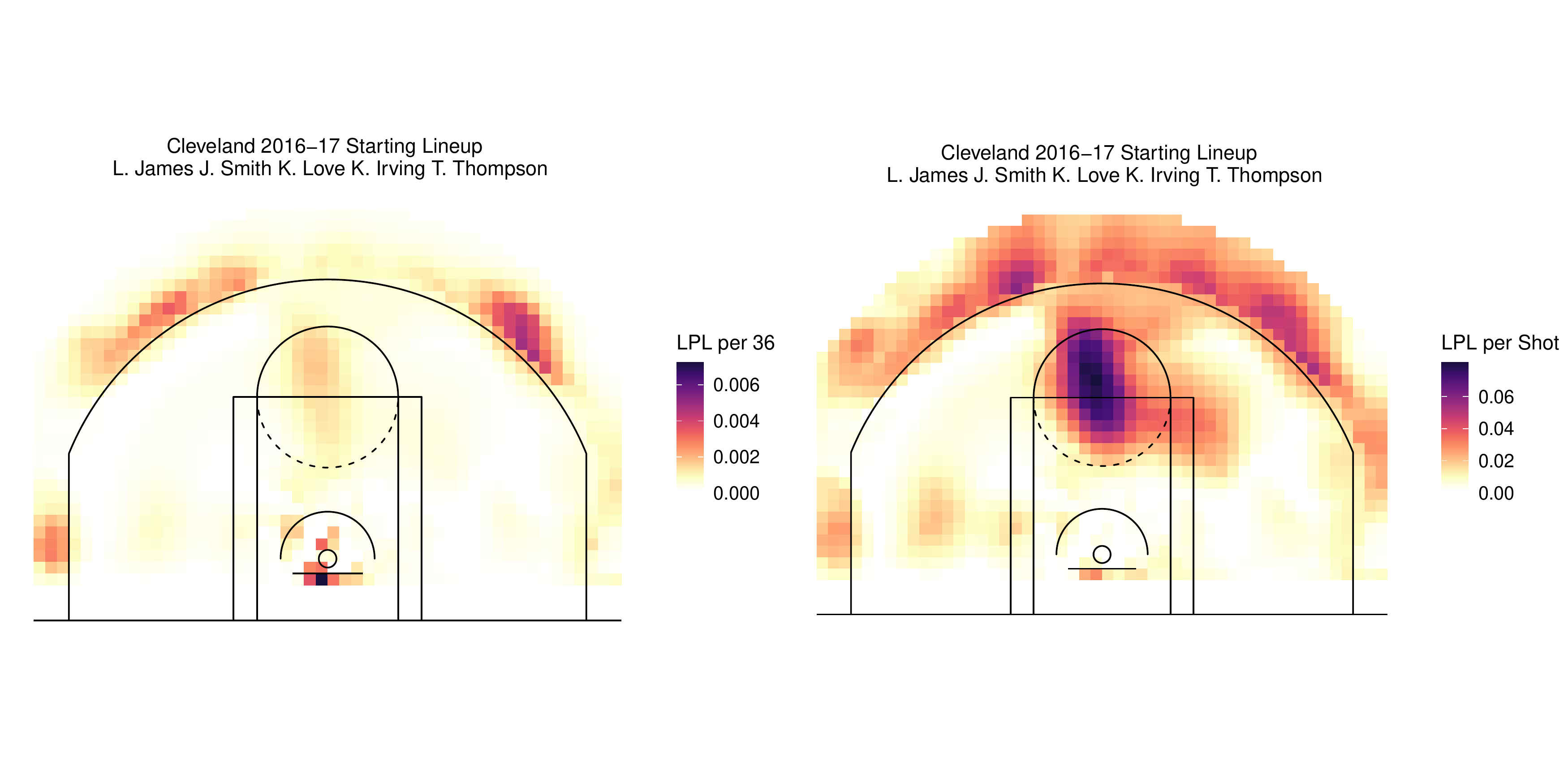}
\caption{$\widehat{\text{LPL}}$ and $\widehat{\text{LPL}}^{Shot}$ surfaces for the Cleveland Cavaliers starting lineup.}
\label{fig:example_LPL}
\end{figure}

The left plot of Figure \ref{fig:example_LPL} shows $\widehat{\text{LPL}}$ (per 36 minutes) over the offensive half court for Cleveland's starting lineup, computed using the posterior mean of $\pmb{\xi}$.\footnote{Since $\text{LPL}_i$ is a function of $\pmb{\xi}_i$, which is latent, the uncertainty in $\text{LPL}_i$ is proportional to the posterior distribution of $\sum_{j = 1}^5 \xi_{ij}$.  Figures \ref{fig:CLE_lpl_distribution}-\ref{fig:LPL_uncertainty_CLE_1} in the appendix illustrate the distributional nature of LPL.}  Notice that the LPL values are highest around the rim and along the 3-point line.  These regions tend to dominate LPL values because the density of shot attempts is highest in these areas. 
 If we re-normalize LPL with respect to the number of shots taken in each court cell we can identify areas of inefficiency that do not stand out due to low densities of shot attempts:  
  \begin{align}
     \text{LPL}_i^{Shot} &= \frac{\text{LPL}_i}{\sum_{j = 1}^5 A_{ij}}.
      \label{eq:lpl_per_shot}
 \end{align}
 This formulation yields the average lineup points lost per shot from region $i$, as shown in the right plot of Figure \ref{fig:example_LPL}.

LPL incorporates an intentional constraint---for any court cell $i$,  $\pmb{A}^*_{i}$ is constrained to be a \textit{permutation} of $\pmb{A}_{i}$.  This ensures that no single player can be reallocated every shot that was taken by the lineup (unless a single player took all of the shots from that region to begin with).  It also ensures that the total number of shots in the redistribution will always equal the observed number of shots from that location \big(i.e. $\sum_{j = 1}^5 A_{ij} = \sum_{j = 1}^5 A^*_{ij}$, for all $i$\big). 

  Ultimately, LPL aims to quantify the points that could have been gained had a lineup adhered to the shot allocation strategy defined by $\pmb{A}^*$.  However, as will be detailed in Section \ref{sec:optimality}, there is not a 1-to-1 relationship between `lineup points' as defined here, and actual points.  In other words, reducing the total LPL of a team's lineup by 1 doesn't necessarily correspond to a 1-point gain in their actual score.  In fact, we find that a 1-point reduction in LPL corresponds to a 0.6-point gain (on average) in a team's actual score.  One reason for this discrepancy could be because LPL is influenced by contextual variables that we are unable to account for in our FG\% model, such as the shot clock and defensive pressure. Another may be due to a tacit assumption in our definition of LPL.  By holding each player's FG\% constant despite changing their volume of shots when redistributing the vector of FG attempts, we implicitly assume that a player's FG\% is independent of their FGA rate.  The basketball analytics community generally agrees that this assumption does not hold---that the more shots a player is allocated, the less efficient their shots become.  This concept, referred to as the `usage-curve' or `skill-curve', was introduced in \cite{oliver2004basketball} and has been further examined in \cite{goldman2011}.  Incorporating usage curves into LPL could be a promising area of future work.  

\subsection{Player LPL Contribution}
 LPL summarizes information from all players in a lineup into a single surface, compromising our ability to identify how each individual player contributes to LPL.  Fortunately, we can parse out each player's contribution to LPL and distinguish between points lost due to undershooting and points lost due to overshooting.  We define player $j$'s LPL contribution (PLC) in court location $i$ as
\begin{align}
  \text{PLC}_{ij} &= \text{LPL}_{i} \times \Bigg(\frac{A^*_{ij} - A_{ij}}{\sum_{j = 1}^5 |A^*_{ij} - A_{ij}|}\Bigg),
  \label{eq:plc}
\end{align}
where all terms are as defined in the previous section.  The parenthetical term in (\ref{eq:plc}) apportions $\text{LPL}_{i}$ among the 5 players in the lineup proportional to the size of their individual contributions to $\text{LPL}_{i}$.  Players who are reallocated more shots under $\pmb{A}^*_{i}$ compared to their observed number of shot attempts will have $\text{PLC}_{ij} > 0$.  Therefore, positive PLC values indicate potential undershooting and negative values indicate potential overshooting.  As in the case of LPL, if we divide PLC by the sum of shot attempts in cell $i$, we obtain average PLC per shot from location $i$:
  \begin{align}
     \text{PLC}_i^{Shot} &= \frac{\text{PLC}_i}{\sum_{j = 1}^5 A_{ij}}.
      \label{eq:plc_per_shot}
 \end{align}
 
\begin{figure}[H]
 \centering
\includegraphics[trim={0cm 3.8cm 0cm 3.8cm}, clip, width=1\textwidth]{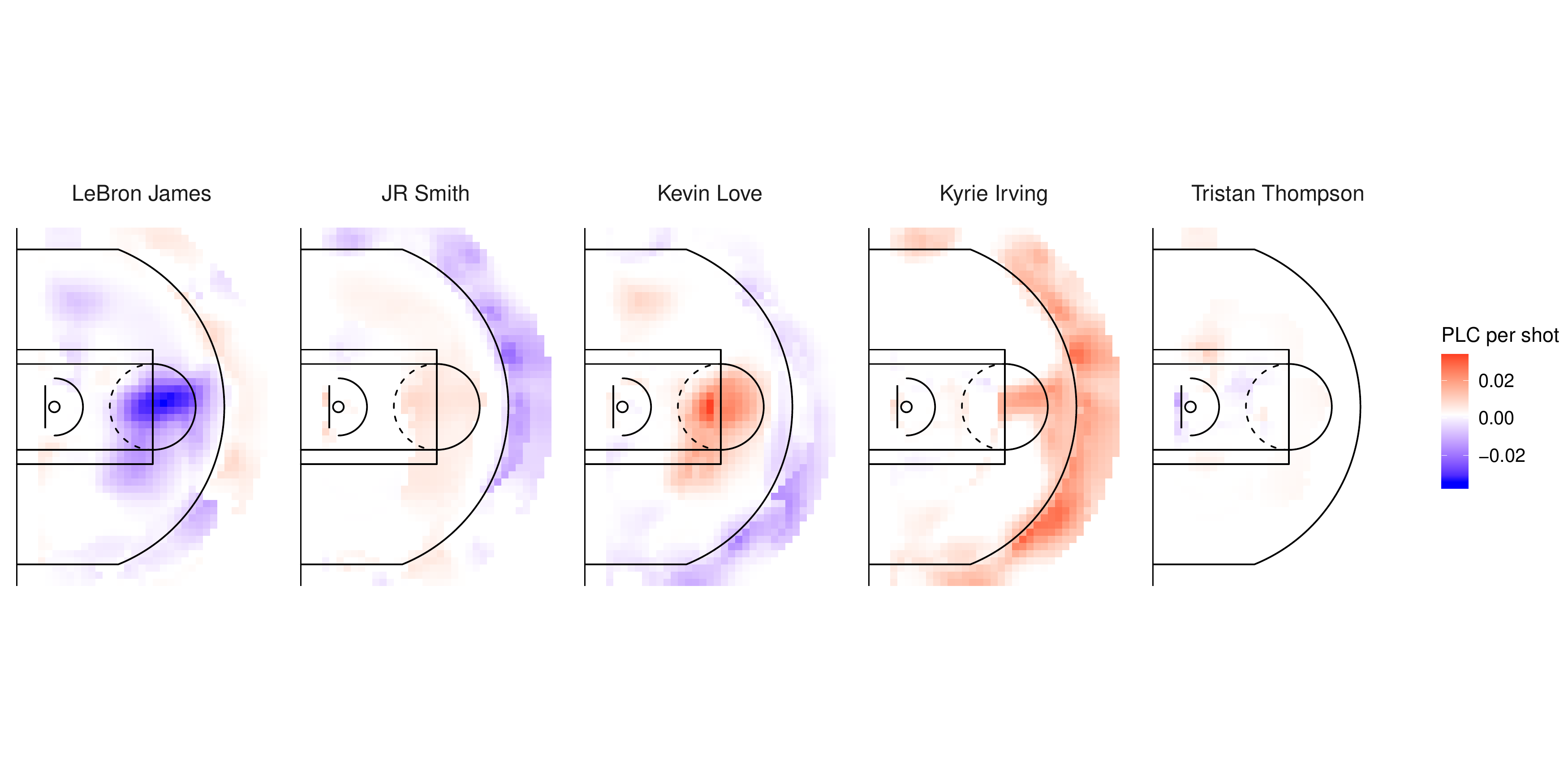}
\caption{$\widehat{\text{PLC}}^{Shot}$ surfaces for the Cleveland Cavaliers starting lineup.}
\label{fig:example_PLC}
\end{figure}

 The $\text{PLC}_i^{Shot}$ surfaces for the Cleveland Cavaliers' 2016-17 starting lineup are shown in Figure \ref{fig:example_PLC}.  We see that Kyrie Irving is potentially being under-utilized from beyond the arc and that LeBron James is potentially over-shooting from the top of the key, which is harmonious with our observations from Figure \ref{fig:example_rank_corr}.  However, it is worth noting that the LPL per 36 plot (left plot in Figure \ref{fig:example_LPL}) shows very low LPL values from the mid-range region since the Cavaliers have a very low density of shots from this area.  So while it may be true that LeBron tends to overshoot from the top of the key relative to his teammates, the lineup shoots so infrequently from this area that the inefficiency is negligible. 

For every red region in Figure \ref{fig:example_PLC} (undershooting) there are corresponding blue regions (overshooting) among the other players.  This highlights the fact that LPL is made up of balancing player contributions from undershooting and overshooting; for every player who overshoots, another player (or combination of players) undershoots.  By nature of how LPL is constructed, there cannot be any areas where the entire lineup overshoots or undershoots.  For this reason, our method does not shed light on shot selection. LPL and PLC say nothing about whether shots from a given region are efficient or not, instead they measure how efficiently a lineup adheres to optimal allocative efficiency given the shot attempts from that region.

\section{Optimality - Discussion and Implications} \label{sec:optimality}

We have now defined LPL and given the theoretical interpretation (i.e. overuse and underuse), but we have not yet established that this interpretation is valid in practice.  The utility of LPL as a diagnostic tool hinges on the answers to four questions, which we explore in detail in this section:
\begin{center}
\begin{quote}
\normalsize{
     1.  Do lineups minimize LPL? \\
     2.  Does LPL relate to offensive production? \\
     3.  How can LPL inform strategy?\\
     4.  Is minimizing LPL always optimal?
     }
\end{quote}
\end{center}

\subsection{Do lineups minimize LPL?} \label{sec:minimize}

In Figure \ref{fig:example_LPL}, cell values range from 0 to 0.008, and the sum over all locations in the half court is 0.68. While this suggests that the Cavaliers' starters were minimizing LPL, we need a frame of reference to make this claim with certainty.  The frame of reference we will use for comparison is the distribution of LPL under completely random shot allocation.   This is not to suggest offenses select shooting strategies randomly.  Rather, a primary reason why lineups fail to effectively minimize LPL is because the defense has the opposite goal; defenses want to get the opposing lineup to take shots from places they are bad at shooting from.  In other words, while the offense is trying to minimize LPL, the defense is trying to maximize LPL.  By comparing LPL against random allocation, this provides a general test for whether offenses are able to pull closer to the minimum than defenses are able to pull toward the maximum, or the absolute worst allocation possible.

In statistical terms, this comparison can be stated as a hypothesis test.  We are interested in testing the null hypothesis that offenses minimize and defenses maximize LPL with equal magnitudes.  We consider a one-sided alternative---that the offensive minimization outweighs the defensive response (as measured on by LPL).  A permutation test allows us to test these hypotheses by comparing a lineup's observed total LPL (summing over all court locations, $\sum_i^M \textnormal{LPL}_i$, where $M$ is the total number of 1 ft. by 1 ft. cells in the half court) against the total LPL we would expect under completely random shot allocation.  To ensure the uncertainty in $\pmb{\xi}$ is accounted for, we simulate variates of the test statistic $T$ as
  \begin{align}
 T &= \sum_{i=1}^M \widetilde{\textnormal{LPL}}_{i=1}^{H_0} - \sum_{i=1}^M \widetilde{\textnormal{LPL}}_i \label{eq:lpl_random1} \\ 
 &= \Bigg(\sum_{i=1}^M \sum_{j = 1}^5 \text{v}_i \cdot \widetilde{\xi}_{ij} \cdot \big(A^*_{ij} - A^{\dagger}_{ij}\big)\Bigg) - \Bigg(\sum_{i=1}^M \sum_{j = 1}^5 \text{v}_i \cdot \widetilde{\xi}_{ij} \cdot \big(A^*_{ij} - A_{ij}\big)\Bigg) \label{eq:lpl_random2}  \\
 &= \sum_{i=1}^M \sum_{j = 1}^5 \text{v}_i \cdot \widetilde{\xi}_{ij} \cdot \big(A_{ij} - A^{\dagger}_{ij}\big), \label{eq:lpl_random3} 
 \end{align}
where $\widetilde{\xi}_{ij}$ is a sample from player $j$'s posterior distribution of FG\% in cell $i$,  $A^{\dagger}_{ij}$ is the $j$th element of a random permutation of the observed FGA rate vector $\pmb{A}_{i}$, and all other symbols are defined as in (\ref{eq:lpl1})-(\ref{eq:lpl2}).  Note that a \textit{different} random permutation is drawn for each court cell $i$.  After simulating 500 variates from the null distribution, we approximate the one-sided p-value of the test as the proportion of variates that are less than 0.  

\begin{figure}[H]
 \centering
\includegraphics[trim={0cm 1.2cm 0cm .8cm}, clip, width=.75\textwidth]{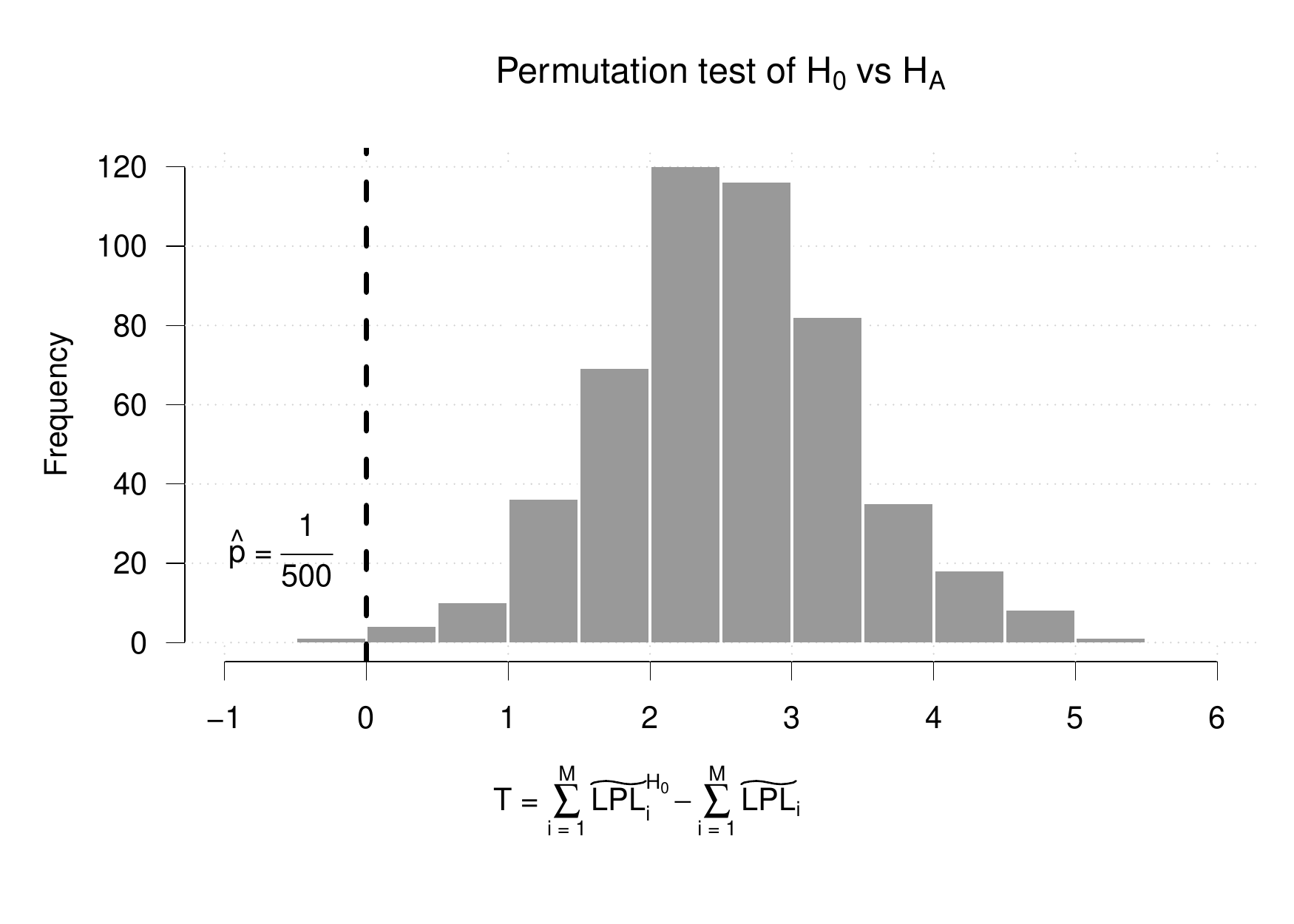}
\caption{Permutation test for the Cleveland Cavaliers' 2016-17 starting lineup.  The gray bars show a histogram of the variates from (\ref{eq:lpl_random1}).  The approximate p-value for the Cavaliers starting lineup (i.e. the proportion of variates that are less than 0) is 1/500 or 0.002.}
\label{fig:lpl_permutation_test}
\end{figure}

Figure \ref{fig:lpl_permutation_test} illustrates this test for the Cleveland Cavaliers' starting lineup.  The gray bars show a histogram of the variates from (\ref{eq:lpl_random1}).  Bars to the left of the dashed line at 0 represent variates for which random allocation outperforms the observed allocation.  The approximate p-value of the test in this case is 1/500, or 0.002.  We can therefore say with high certainty that the Cleveland starters minimize LPL beyond the defense's ability to prevent them from doing so.   
The computational burden of performing the test precludes performing it for every lineup, but we did perform the test for each team's 2016-17 starting lineup.  The results are shown in Table \ref{tab:p-value}.  Across the NBA's starting lineups, only two teams had no variates less than 0---the Golden State Warriors and the Portland Trailblazers.  The Sacramento Kings showed the worst allocative efficiency with an approximate p-value of 0.44 for their starting lineup.  Based on these results we are confident that most lineups employ shot allocation strategies that minimize LPL to some degree, though it appears that some teams do so better than others.
\begin{table}[ht]
\small
\centering
\textbf{Approximate p-values for} $\text{H}^0$ \textbf{vs.} $\text{H}^{\text{A}}$
%\begin{tabular}{rllllllllllllllllllllllllllllll}
\resizebox{\textwidth}{!}{\begin{tabular}{rlllllllllllllll}
  \hline
 & 1 & 2 & 3 & 4 & 5 & 6 & 7 & 8 & 9 & 10 & 11 & 12 & 13 & 14 & 15 \\
  \hline
Team & GSW & POR & CLE & LAC & ATL & HOU & TOR & IND & LAL & DET & DEN & NOP & CHA & UTA & OKC \\ 
  $\hat{p}$ & 0.000 & 0.000 & 0.002 & 0.002 & 0.014 & 0.014 & 0.016 & 0.020 & 0.022 & 0.024 & 0.028 & 0.030 & 0.030 & 0.038 & 0.042 \medskip \\
    \hline
   & 16 & 17 & 18 & 19 & 20 & 21 & 22 & 23 & 24 & 25 & 26 & 27 & 28 & 29 & 30 \\
   \hline
Team & DAL & MIA & MIN & BOS & NYK & ORL & SAS & BKN & PHI & MIL & WAS & PHX & MEM & CHI & SAC \\ 
 $\hat{p}$ & 0.044 & 0.046 & 0.054 & 0.056 & 0.058 & 0.064 & 0.104 & 0.106 & 0.130 & 0.134 & 0.144 & 0.148 & 0.170 & 0.210 & 0.442 \\ 
   \hline
%\end{tabular}
\end{tabular}}
\caption{Approximate p-values for $\text{H}^0$ vs. $\text{H}^{\text{A}}$ for each team's starting lineup in the 2016-17 NBA regular season.}
\end{table} \label{tab:p-value}

\subsection{Does LPL relate to offensive production?} \label{sec:dixon_coles}

We next want to determine whether teams with lower LPL values tend to be more proficient on offense.  In order to achieve greater discriminatory power, we've chosen to make this assessment at the game level.  Specifically, we regress a team's total game score against their total LPL generated in that game, accounting for other relevant covariates including the team's offensive strength, the opponents' defensive strength, and home-court advantage.  This framework is analogous to the model proposed in \cite{dixon1997modelling}. 

We calculate game LPL (GLPL) by first dividing the court into three broad court regions (restricted area, mid-range, and 3-pointers).  Then, for a given game and lineup, we calculate GLPL in each of these court regions (indexed by $c$) by redistributing the lineup's observed vector of shot attempts using on a weighted average of each player's $\widehat{\pmb{\xi}}_j$:
  \begin{align}
  \text{GLPL}_c &= \sum_{j = 1}^5 \text{v}_c \cdot f_c(\widehat{\pmb{\xi}}_{j}) \cdot \big(A^*_{cj} - A_{cj}\big), ~~~ \text{where} ~~~ f_c(\widehat{\pmb{\xi}}_{j}) = \frac{\sum_{i \in c}w_{ij}\widehat{\xi}_{ij}}{\sum_{i \in c}w_{ij}}. \label{eq:fg_weight} 
  \end{align}
In (\ref{eq:fg_weight}), $w_{ij}$ is a weight proportional to player $j$'s total observed shot attempts in court cell $i$ over the regular season.  The notation $\sum_{i \in c}$  means we are summing over all the 1 ft. by 1 ft. grid cells that are contained in court region $c$.  Finally, for a given game $g$ and team $a$, we calculate the team's total game LPL (TGLPL) by summing $\text{GLPL}_c$ over all court regions $c$ and all lineups $\ell$:
  \begin{align}
  \text{TGLPL}_{ag} &= \sum_{\ell = 1}^{L_a} \sum_{c \in C} \text{GLPL}_c^{\ell}  \label{eq:TGL} 
 \end{align}
 where $C = \{\text{restricted area, mid-range, 3-pointers\}}$ and $L_a$ is the total number of team $a$'s lineups.  This process is carried out separately for the home and away teams, yielding two TGLPL observations per game. 
 
Equipped with a game-level covariate measuring aggregate LPL, we model team $a$'s game score against opponent $b$ in game $g$ as
\begin{align}
    \text{Score}_{abg} &= \mu + \alpha_a + \beta_b + \gamma \times \text{I}(\text{Home}_{ag}) + \theta \times \text{TGLPL}_{ag} + \epsilon_{abg} \label{eq:dixon_coles1}\\
    \epsilon_{abg} &\sim N(0, \sigma^2), \label{eq:dixon_coles2}
\end{align}
where $\mu$ represents the global average game score, $\alpha_a$ is team $a$'s offensive strength parameter, $\beta_b$ is team $b$'s defensive strength parameter, $\gamma$ governs home court advantage, $\theta$ is the effect of TGLPL, and $\epsilon_{abg}$ is a normally distributed error term.  $\theta$ is the parameter that we are primarily concerned with.  We fit this model in a Bayesian framework using Hamiltonian Monte Carlo methods implemented in Stan \citep{carpenter2017stan}.  Our prior distributions are as follows: $\mu \sim N(100, 10^2)$; $\alpha_a, \beta_b, \gamma, \theta \sim N(0, 10^2)$; $\sigma \sim Gamma(\text{shape} = 2, \text{rate} = 0.2)$. 

\begin{figure}[H]
 \centering
\includegraphics[trim={0cm 0cm 0cm 0cm}, clip, width=1\textwidth]{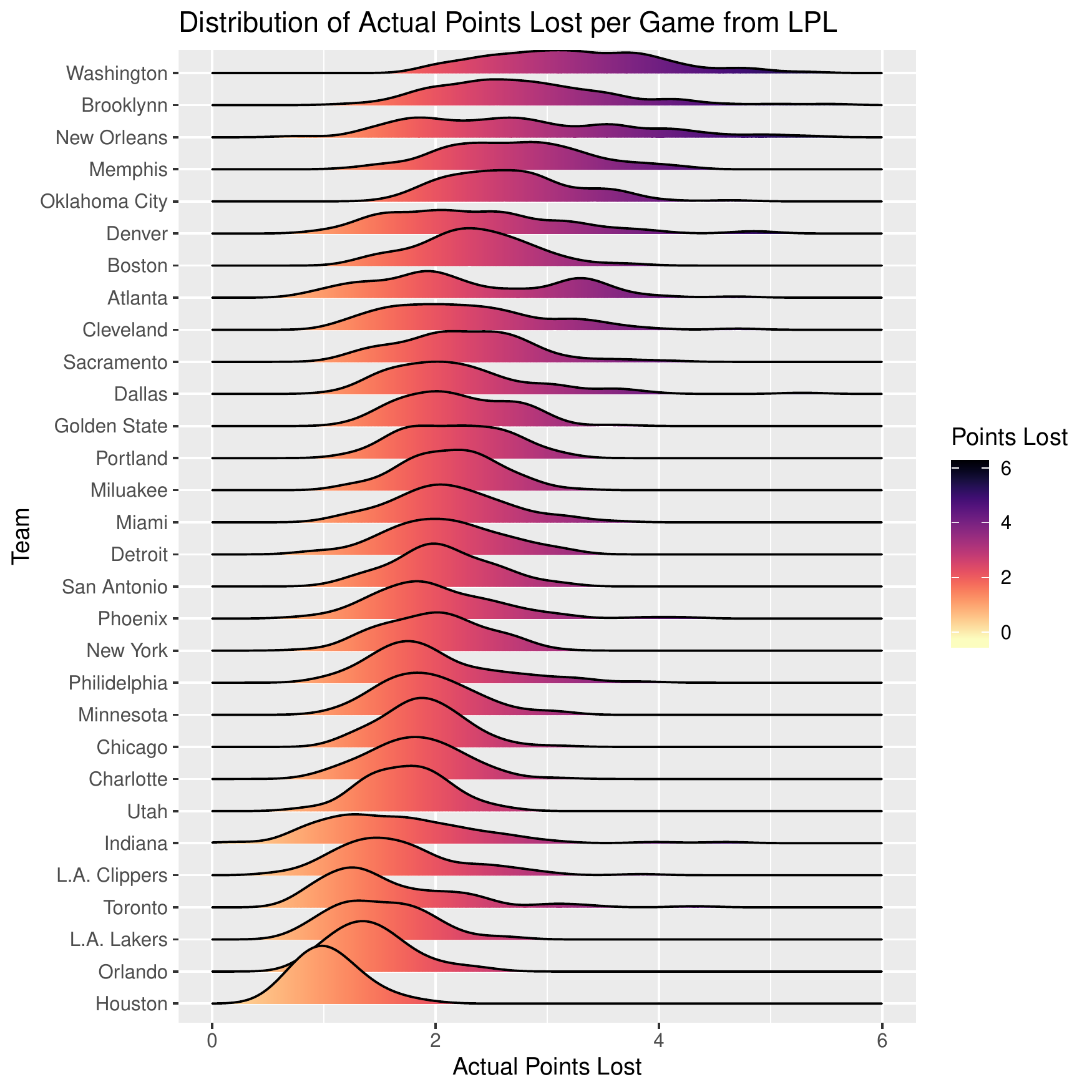}
\caption{Estimated density of actual points lost per game for every team's 82 games in the 2016-17 NBA regular season.}
\label{fig:game_lpl_team}
\end{figure}

The 95\% highest posterior density interval for $\theta$ is (-1.08, -0.17) and the posterior mean is -0.62.\footnote{Figure \ref{fig:theta_posterior} in the appendix shows the posterior distribution of $\theta$.}  Therefore, we estimate that for each additional lineup point lost, a team loses 0.62 actual points.  Put differently, by shaving roughly 3 points off of their TGLPL, a team could gain an estimated 2 points in a game.  Given that 10\% of games were decided by 2 points or less in the 2016-17 season, this could have a significant impact on a team's win-loss record and could even have playoff implications for teams on the bubble.  Figure \ref{fig:game_lpl_team} shows the estimated density of actual points lost per game for every team's 82 games in the 2016-17 NBA regular season (i.e. density of $\widehat{\theta} \times \text{TGLPL}_{ag},~ g \in \{1,\ldots, 82\} \text{ for each team } a)$.  Houston was the most efficient team, only losing about 1 point per game on average due to inefficient shot allocation.  Washington, on the other hand, lost over 3 points per game on average from inefficient shot allocation.  

\subsection{How can LPL inform strategy?}

At this point, we offer some ideas for how coaches might use these methods to improve their teams' offense.  First, for lineups with high LPL, coaches could explore the corresponding PLC plots to ascertain which players are primarily responsible.  If the coach determines that the LPL values do indeed represent areas of inefficiency, they could consider interventions targeting the player's shooting habits in these areas.  This short-term intervention could be coupled with long-term changes to their practice routines; coaches could work with players on improving their FG\% in the areas shown by the PLC plots.  Also, by exploring lineup PLC charts, coaches could identify systematic inefficiency in their offensive schemes, which could prompt changes either in whom to draw plays for or whether to change certain play designs altogether.  

Coaches are not the only parties who could gain value from these metrics; players and front office personnel could utilize them as well.  Players could use PLC plots to evaluate their shooting habits and assess whether they exhibit over-confident or under-confident shot-taking behavior from certain areas of the court.  Front office personnel may find trends in the metrics that indicate a need to sign players that better fit their coach's strategy.  LPL and PLC could help them identify which players on their roster to shop and which players to pursue in free agency or the trade market.  

\begin{figure}[H]
 \centering
 \includegraphics[trim={0cm 3.5cm 0cm 3.2cm}, clip, width=1\textwidth]{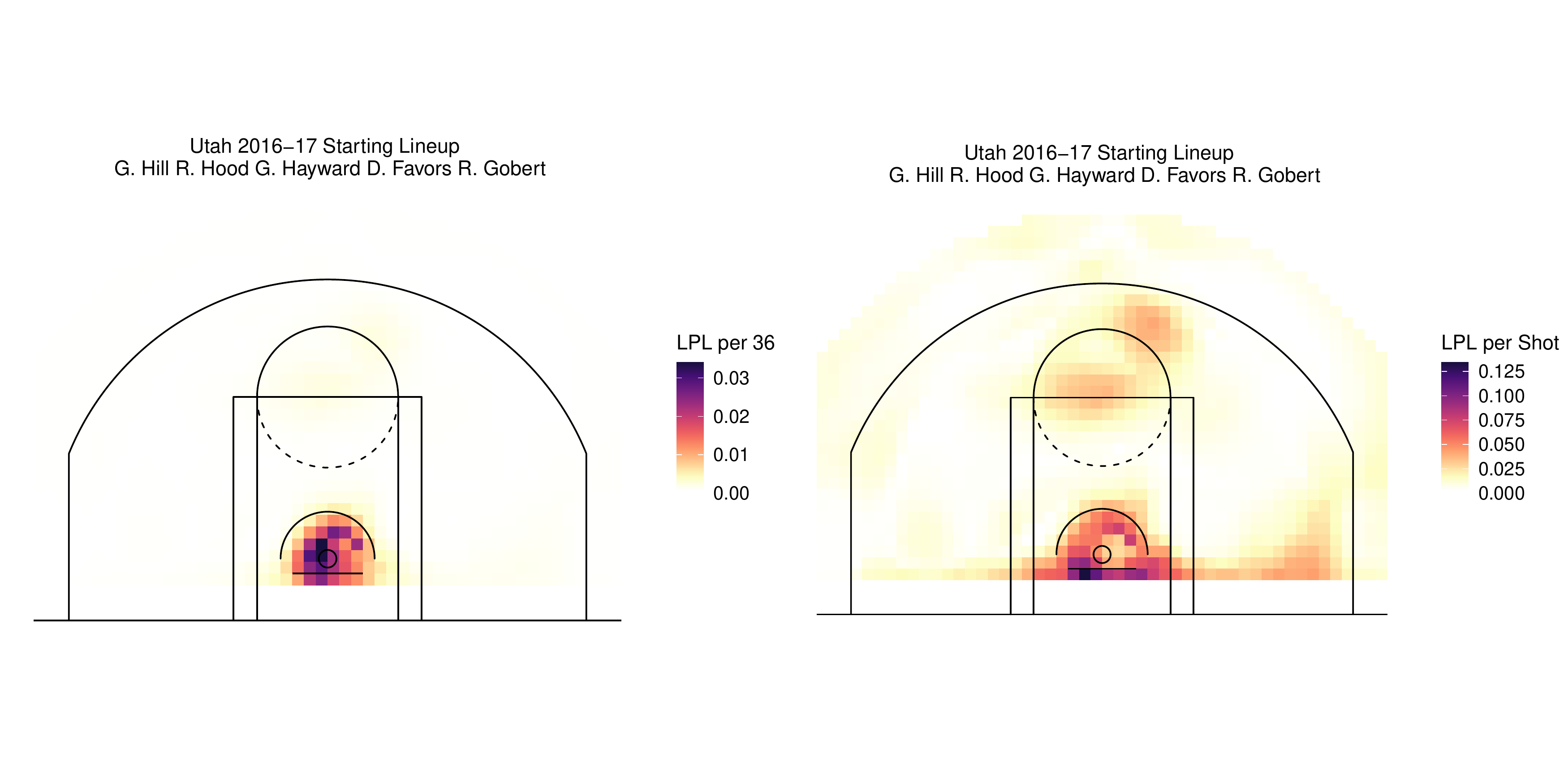}
\includegraphics[trim={0cm 4.6cm 0cm 3.4cm}, clip, width=1\textwidth]{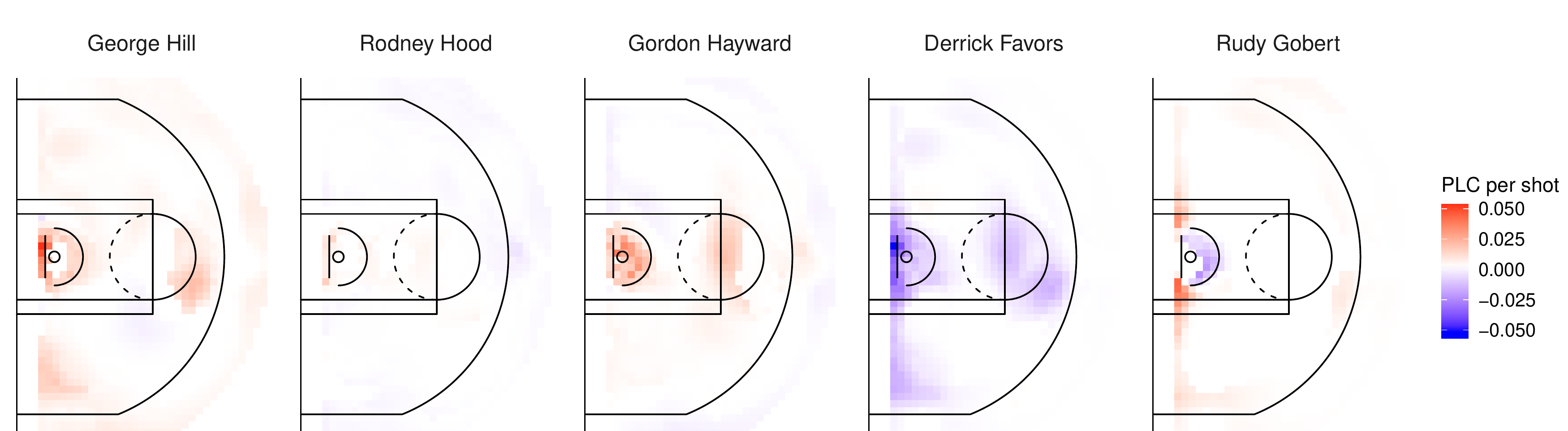}
\caption{Utah Jazz 2016-17 starting lineup $\widehat{\text{LPL}}$, $\widehat{\text{LPL}}^{Shot}$, and $\widehat{\text{PLC}}^{Shot}$ surfaces.}
\label{fig:UTA_PLC}
\end{figure}

Consider these ideas in context of the Utah Jazz LPL/PLC charts for the 2016-17 regular season shown in Figure \ref{fig:UTA_PLC}.
On reviewing the LPL per shot plot for the starting lineup, the coach might flag the left baseline and top of the key as areas of potential inefficiency to investigate.  On exploring the corresponding PLC plots, they would see Derrick Favors as the driving force behind the high LPL numbers from these regions.  Interestingly, from the 2013-14 season through 2016-17, the Derrick Favors baseline and elbow jump shots were go-to plays for the Jazz.  Across these four seasons, Favors took over 1500 mid-range shots for an average of 0.76 points per shot (PPS). 

In the 2017-18 and 2018-19 seasons, the Jazz drastically altered Favors' shot policy from the mid-range.  Beginning in 2017, the Jazz started focusing on running plays for 3-pointers and shots at the rim, a trend that was becoming popular throughout the league.  As part of this change in play-style, they tried turning Favors into a stretch four\footnote{A stretch four is a player at the power forward position that can generate offense farther from the basket than a traditional power forward.}; he went from taking a total of 21 3-point shots over the previous four seasons, to 141 3-point shots in these two seasons alone.  Unfortunately, their intervention for Favors appears to have been misguided; his average PPS for these 141 shots was 0.66.  The front office eventually determined that Favors wasn't the best fit for their coach's offensive strategy; they opted not to re-sign Favors at the end of the 2019 season. 

This process took place over six years—perhaps it could have been expedited had LPL and PLC been available to the coaches and front office staff.

\subsection{Is minimizing LPL always optimal?}
While we have demonstrated that lower LPL is associated with increased offensive production, we stress that LPL is a diagnostic tool that should be used to inform basketball experts rather than as a prescriptive measure that should be strictly adhered to in all circumstances.  As mentioned previously, the LPL and PLC values presented in this paper are influenced by contextual variables that we are unable to account for because they are not available in public data sources, such as the shot clock and defensive pressure.  Additionally, there are certain game situations where minimizing LPL may be sub-optimal.  

\begin{figure}[H]
 \centering
\includegraphics[trim={0cm 4.5cm 0cm 4.2cm}, clip, width=1\textwidth]{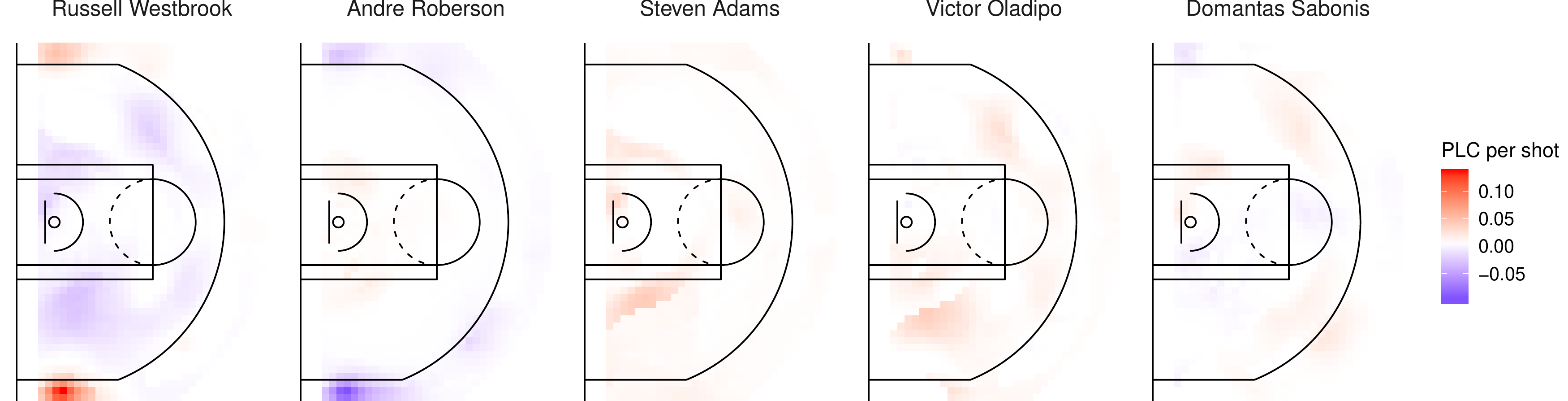}
\caption{Oklahoma City 2016-17 starting lineup $\widehat{\text{PLC}}^{Shot}$ surfaces.}
\label{fig:OKC_PLC}
\end{figure}

One such situation is illustrated in Figure \ref{fig:OKC_PLC}, which shows the $\text{PLC}^{Shot}$ surfaces for the Oklahoma City 2016-17 starting lineup. 
The first panel from the left in this figure shows positive PLC values for Russell Westbrook in the corner 3-point regions, suggesting that Westbrook should be taking more shots from these areas.  However, anyone who watched the Thunder play that season will know that many of these corner 3-point opportunities were created by Westbrook driving to the basket, drawing extra defenders toward him, then kicking the ball out to an open teammate in the corner.  Obviously, Westbrook cannot both drive to the rim and simultaneously pass to himself in another area of the court.  In this case, strictly minimizing LPL would reduce the number of these drive-and-kick plays, potentially attenuating their offensive firepower.  Shot-creation is not accounted for by LPL and should be carefully considered when exploring LPL and PLC.  

There are game theoretic factors to be considered as well.  Beyond the defensive elements discussed in Section 4.1, rigid adherence to minimizing LPL could lead to a more predictable offense and thus make it easier to defend \citep{damour2015}.   Needless to say, offensive game-planning should be informed by more than LPL metrics alone.  

 \section{Conclusion}

 Our research introduces novel methods to evaluate allocative efficiency spatially and shows that this efficiency has a real impact on game outcomes.  We use publicly available data and have made an empirical demonstration of our methods available online, allowing our methods to be immediately accessible. Also, since LPL and PLC do not depend on specific models for FG\% and FGA rate, LPL and PLC could readily be calculated at G-league, NCAA, and international levels using a simplified model of FG\% and FGA rate. 
 
As most professional basketball teams have access to proprietary data, many of the contextual variables that we do not account for could be included in the FG\% and FGA rate models, which could make the proposed shot distribution proposed by LPL a more reliable optimum to seek.  Additionally, by pairing LPL with play call data coaches could gain insight into the efficiency of specific plays.  Even without access to these data, it may be possible to recreate some contextual features that aren't explicitly provided by the NBA's public-facing API.  For instance, shot clock times could be reverse engineered using game clock times given in the play-by-play data. 

There are interesting academic questions that stem from this paper as well.  Future studies could investigate the sensitivity of our metrics to model parameters that we fixed, such as the number of basis functions in the NMF and the number of neighbors in the CAR prior.  We could also investigate the robustness of LPL to alternate FG\% models.  As mentioned previously, we do not account for usage curves in our analysis.  Doing so would turn LPL into a constrained optimization problem, which would be a fascinating challenge to tackle.  Also, using LPL to inform player-specific shot policy changes, entire seasons could be simulated using the method in \cite{sandholtz2020transition} to quantify the impact of specific shot allocation changes on point production.  We hope that the methods introduced in this paper will be built upon and improved.  

\newpage
\section{Appendix}

\subsection{Empirical Implementation} \label{sec:empirical_example}

To illustrate some important considerations associated with this approach, we present a brief example of LPL and PLC using empirical FG\% and FGA rates. This example demonstrates that these quantities are agnostic to the underlying FG\% model. 

We examine the same lineup for the Cavaliers that is discussed in the main text. In order to obtain FG\% and FGA rate estimates, we divide the court into twelve discrete regions and calculate the empirical values for each player within these regions. We defined these regions based on our understanding of the court, but it is worth noting that defining these regions requires many of the same considerations as with any histogram style estimator; namely, that increasing the number of regions will decrease bias at the expense of increasing variance. In some cases, a player may have only one or two shots within an area, resulting in either unrealistically high or low field goal percentage estimates. As an \textit{ad hoc} solution to this, we give all players one made field goal and five field goal attempts within each region, which means that players with just a handful of shots in a region will have their associated field goal percentage anchored near 20 percent. Rather than perform smoothing for the field goal attempt estimates, we simply count up the number of attempts for each player within each section, and normalize them to get the attempts per 36 minutes, as before. With these FG\% and FGA estimates, we can replicate the analysis detailed in Section 3. 

\begin{figure}[H]
    \centering
\includegraphics[trim={0cm .4cm 0cm 0cm}, clip, width=.85\textwidth]{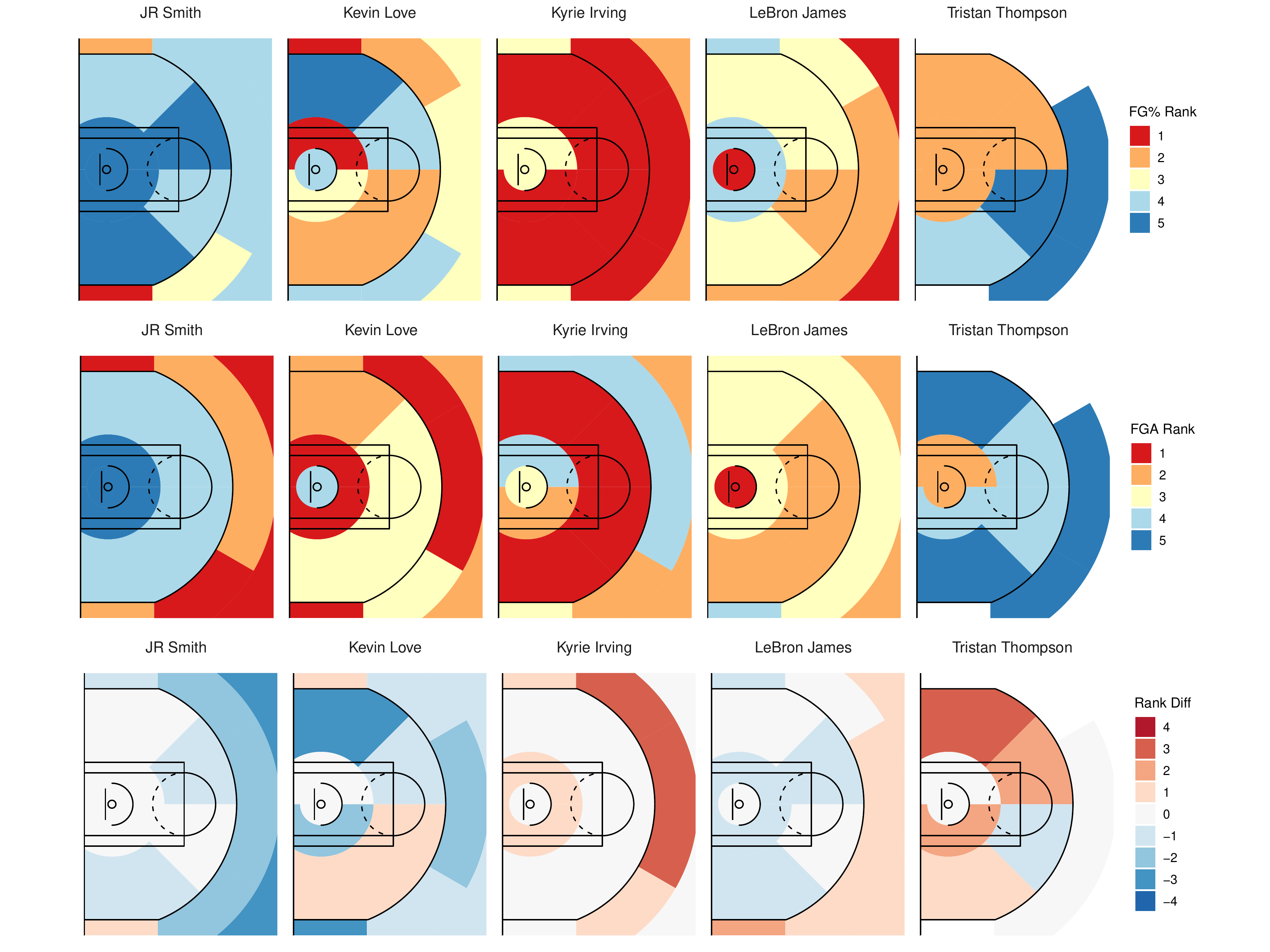}
\caption{Top: Empirical FG\% ranks for the Cleveland Cavaliers starting lineup. Middle: Empirical FGA ranks. Bottom: Rank correspondence.}
\label{fig:empirical_ranks}
\end{figure}

\begin{figure}[H]
    \centering
    \includegraphics[trim={0cm 2.5cm 0cm 2.5cm}, clip, width=.85\textwidth]{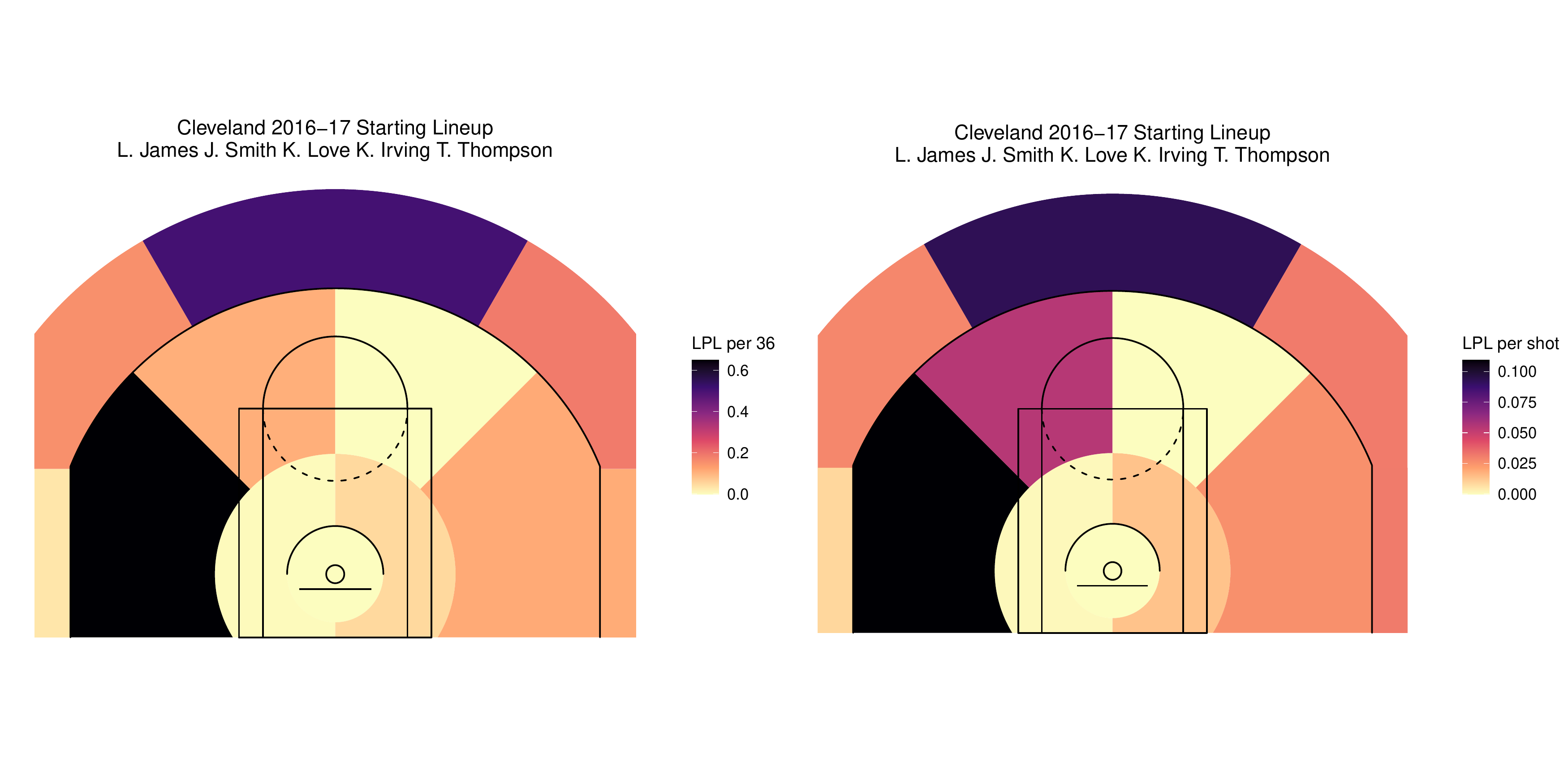}
      \includegraphics[trim={0cm 2.5cm 0cm 2cm}, clip, width=.85\textwidth]{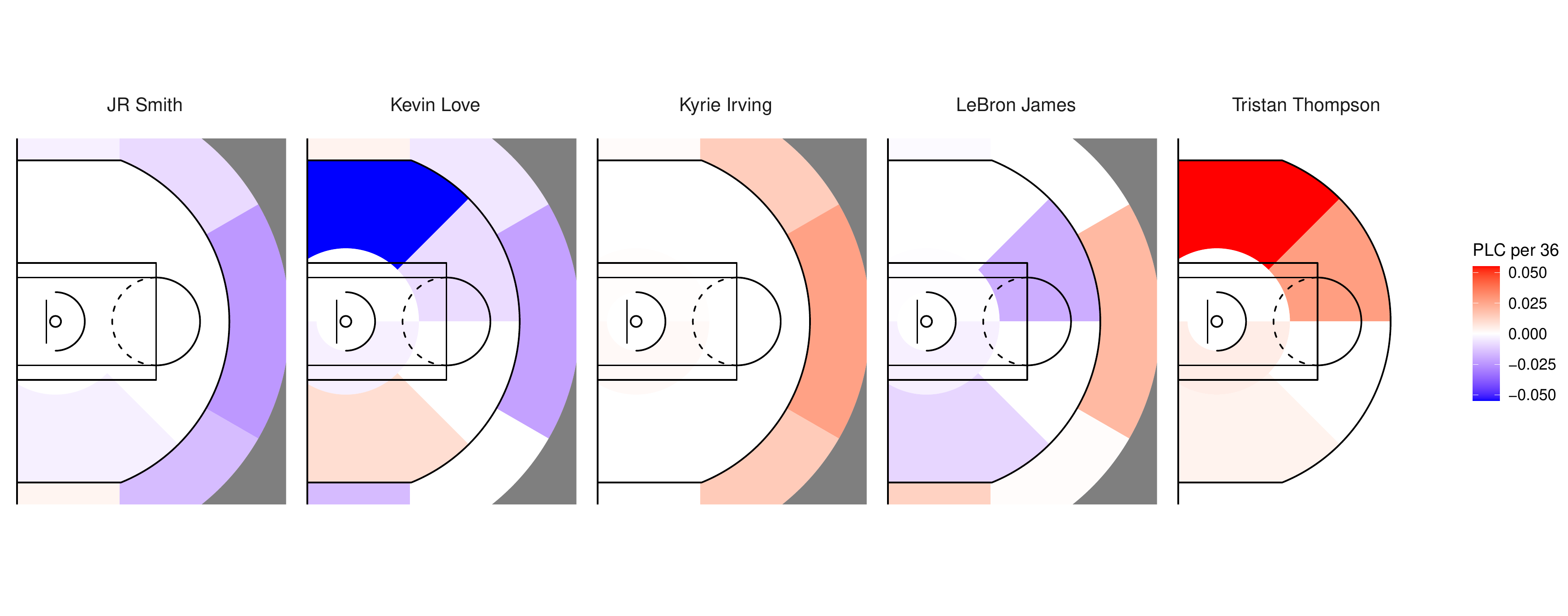}
    \caption{Top: Empirical LPL and $\text{LPL}^{Shot}$ for the Cleveland Cavaliers starting lineup.  Bottom: Empirical PLC for the Cleveland Cavaliers starting lineup.}
    \label{fig:empirical_lpl}
\end{figure}

Figure \ref{fig:empirical_ranks} shows the empirical ranks for this lineup, as well as the rank correspondence. Generally, it shows the same patterns as the model-based analysis in Figures \ref{fig:example_fga_ranks} and \ref{fig:example_rank_corr}. However, there are some key differences, including Tristan Thompson having a higher field goal percentage rank from the right midrange and a corresponding reduction in rank for Kevin Love in the same area. This pattern is also manifest in Figure \ref{fig:empirical_lpl}, which shows the empirical LPL. We observe that most lineup points appear to be lost in the right midrange and in above the break three point shots. Finally, considering the empirical PLC in Figure \ref{fig:empirical_lpl}, we notice that in addition to the Love-Thompson tradeoff in the midrange, JR Smith appears to be overshooting from the perimeter, while Kyrie Irving and LeBron James both exhibit undershooting.

The persistence of the Love-Thompson connection in the midrange in this empirical analysis, and its divergence from what we saw in the model based analysis, merits a brief discussion. Kevin Love and Tristan Thompson both had a low number of shots from the far-right midrange region, with Love shooting 8 for 26 and Thompson shooting 4 for 6. Because they both shot such a low amount of shots, even with the penalty of one make and four misses added to each region, Thompson appears far better. This highlights the fact that although LPL and PLC are model agnostic, the underlying estimates for field goal percentage do matter and raw empirical estimates alone may be too noisy to be useful in calculating LPL. One simple solution may be to use a threshold and only consider players in a region if the number of their field goal attempts passes that threshold.

\subsection{Additional Figures}

\begin{figure}[H]
\begin{center}
\includegraphics[trim={3cm 12.1cm 3cm 0cm}, clip, width=.9\textwidth]{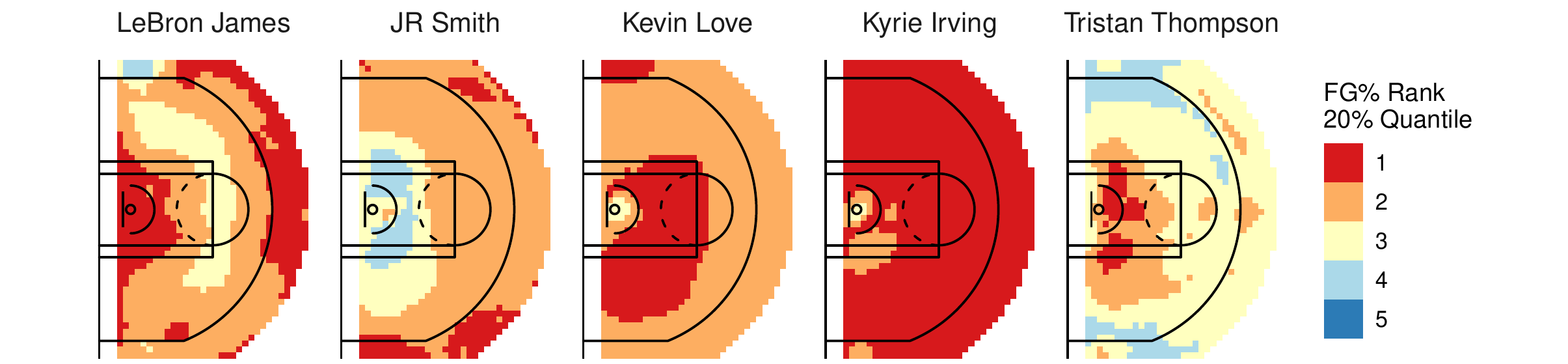}
\includegraphics[trim={3cm 6.25cm 3cm 6.7cm}, clip, width=.9\textwidth]{ranks_low_med_up_CLE_1.pdf}
\includegraphics[trim={3cm 0cm 3cm 12.5cm}, clip, width=.9\textwidth]{ranks_low_med_up_CLE_1.pdf}
\end{center}
\caption{Top: 20\% quantiles of the Cleveland Cavaliers starting lineup posterior distributions of FG\% ranks. Middle: medians of these distributions.  Bottom: 80\% quantiles.}
\label{fig:example_fgp_ranks}
\end{figure}

\begin{figure}[H]
\centering
\includegraphics[trim={0cm 0cm 0cm 0cm}, clip, width=.6\textwidth]{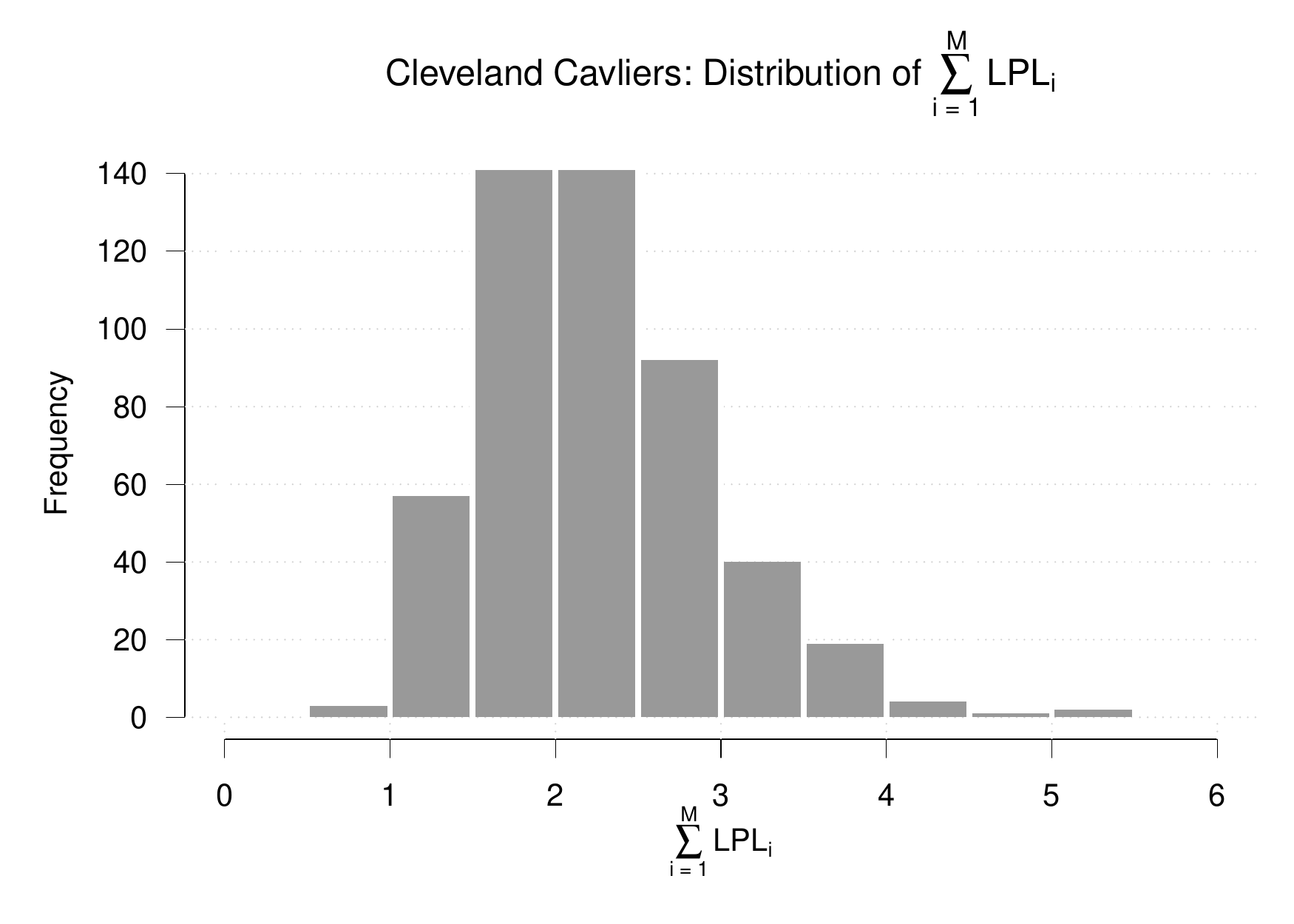}
\caption{Histogram of $\sum_{i=1}^\text{M} \text{LPL}_i$ for the Cleveland Cavaliers starting lineup.  500 posterior draws from each $\xi_{ij}$, where $i \in \{1\ldots,M\} \text{ and } j \in \{1,\ldots,5\}$, were used to compute the 500 variates of $\sum_{i=1}^M \text{LPL}_i$ comprising this histogram.}
\label{fig:CLE_lpl_distribution}
\end{figure}

\vspace{-.25in}
\begin{figure}[H]
\begin{center}
\includegraphics[trim={0cm 0cm 0cm 0cm}, clip, width=1\textwidth]{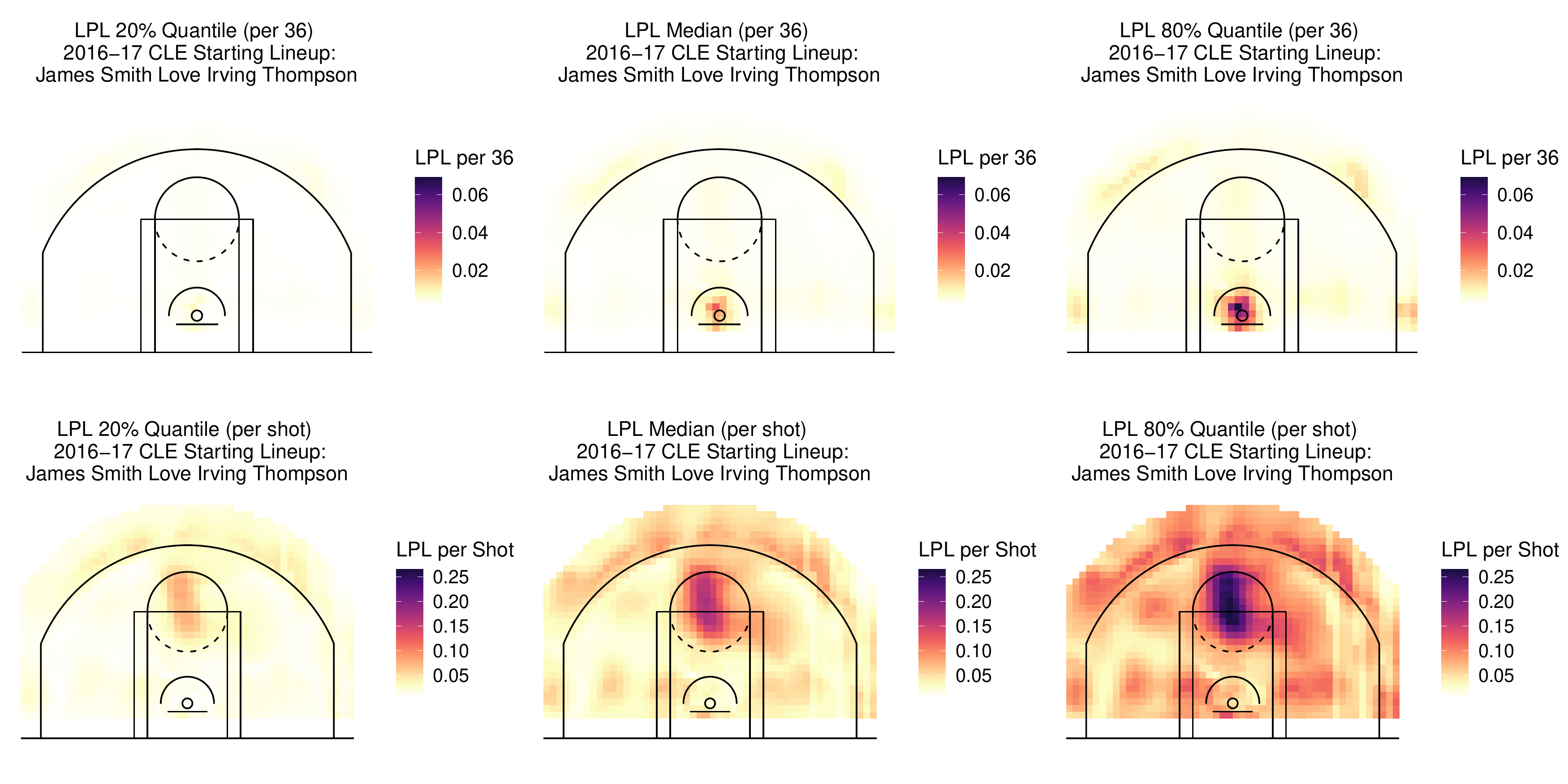}
\end{center}
\caption{Left: 20\% quantile $\text{LPL}$ surfaces for the Cleveland Cavaliers starting lineup. Middle: median $\text{LPL}$ surfaces.  Bottom: 80\% quantile $\text{LPL}$ surfaces.  The top rows show $\text{LPL}^{36}$ while the bottom rows show $\text{LPL}^{\text{Shot}}$.}
\label{fig:LPL_uncertainty_CLE_1}
\end{figure}

\vspace{-.25in}
\begin{figure}[H]
\begin{center}
\includegraphics[trim={0cm .5cm 0cm .5cm}, clip, width=.7\textwidth]{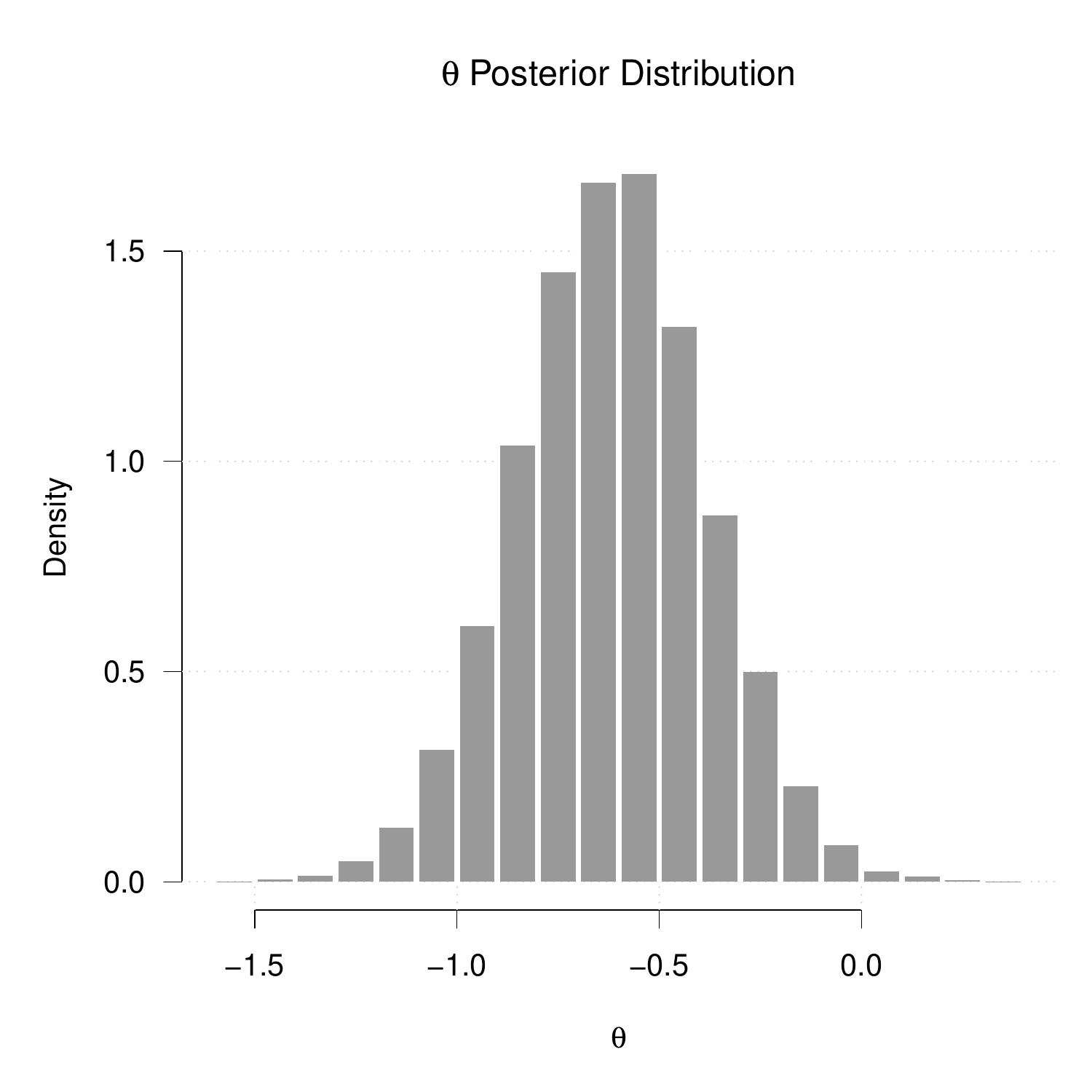}
\end{center}
\caption{Posterior distribution of the effect for TGLPL in model (\ref{eq:dixon_coles1})-(\ref{eq:dixon_coles2}) described in Section \ref{sec:dixon_coles}.}
\label{fig:theta_posterior}
\end{figure}

% \begin{acknowledgement}
%   ...
% \end{acknowledgement}

\bibliographystyle{agsm}
\bibliography{bibliography}
\end{document}